\documentclass{article}

\usepackage{arxiv}

\usepackage[utf8]{inputenc}
\usepackage[T2A,T1]{fontenc}
\DeclareTextFontCommand{\wordcyrillic}{\fontencoding{T2A}\selectfont}
\usepackage{url}        
\usepackage{booktabs}      
\usepackage{array}
\usepackage{amsfonts}     
\usepackage{nicefrac}      
\usepackage{microtype}
\usepackage{lipsum}
\usepackage{graphicx}
\usepackage{tabularx}
\usepackage{float}
\usepackage{placeins}
\usepackage{listings}
\usepackage{caption}
\usepackage{lineno} 

\renewcommand{\figurename}{Fig.}
\DeclareCaptionLabelSeparator{barsep}{\space|\space}
\lstdefinestyle{prompt}{
  basicstyle=\ttfamily\footnotesize,
  columns=fullflexible,
  keepspaces=true,
  showstringspaces=false,
  breaklines=true,
  breakatwhitespace=true,
  frame=single,
  captionpos=t,
  aboveskip=10pt,
  belowskip=10pt,
  xleftmargin=4pt,
  xrightmargin=4pt
}
\usepackage[superscript]{cite}
\usepackage{hyperref}       
\graphicspath{ {./}{./images/} }
\newcommand{\smilescode}[1]{\mbox{\texttt{\scriptsize #1}}}
\title{Physical Reasoning in Large Language Models is Shaped by Human Language}

\author{
  \textbf{Gregorii A. Semakin}$^{1,2,*}$, \quad
  \textbf{Timofey V. Losev}$^{1}$, \quad
  \textbf{Ilya V. Prolomov}$^{1}$, \\
  \textbf{Stepan N. Ostarkov}$^{1}$, \quad
  \textbf{Igor V. Alabugin}$^{3}$, \quad
  \textbf{Michael G. Medvedev}$^{1,*}$ \\[0.8em]
  $^{1}$N. D. Zelinsky Institute of Organic Chemistry RAS, 119991 Moscow, Russia \\
  $^{2}$National Research University Higher School of Economics, 101000 Moscow, Russia \\
  $^{3}$Department of Chemistry and Biochemistry, Florida State University, Tallahassee, FL 32306, USA \\[0.5em]
  \texttt{\{gregoriisemakin@gmail.com, medvedev.m.g@gmail.com\}}
}

\begin{document}
\raggedbottom
\maketitle

\begin{abstract}

Large language models (LLMs) are increasingly used in scientific research \cite{medvedev2026chemistryai,klimesch2026designingphysics,dip2026llmagents}, but whether next-token prediction can yield physical understanding remains unresolved \cite{lecun2022path,klindt2026lejepa,balestriero2026provingimprovingplatonicrepresentation,radford2019language}. Human scientific reasoning relies on verbalizable concepts that can obscure less intuitive effects. Molecular conformations expose this tension: hydrogen bonds offer a recognizable stabilizing pattern (sometimes worth a few kcal/mol\cite{steiner2002hydrogenbond}), whereas strain from subtle ring puckering is harder to quantify. We find the same bias in LLMs: while frontier LLMs are able to rank conformers of diverse organic molecules from atomic coordinates alone approaching accuracy of specialized force-fields, they mostly rely on the better-quantified electrostatic interactions, including hydrogen bonds. They almost universally fail in cases where the hydrogen-bond heuristic points in the wrong direction. Deeper analysis reveals that LLMs do recognize and geometrically characterize less quantified chemical interactions including strain and sterics, but struggle to assess their energetic contributions, just as humans do. These findings suggest that emerging physical reasoning in LLMs inherits both the strengths and blind spots of human scientific concepts—a machine analogue of linguistic relativity \cite{kay1984sapirwhorf}. As researchers increasingly rely on LLMs for scientific judgments, the most capable models produce systematic errors where they are the most dangerous – when human intuition also fails.

\end{abstract}

Large language models (LLMs) operate on discrete sequences of tokens, yet the physical world they are increasingly tasked to understand is continuous and spatial. This contrast lies at the heart of the ongoing debate in artificial intelligence. Yann LeCun advocates world models learned through perception and interaction with the environment as a route to physical understanding\cite{lecun2022path,klindt2026lejepa,balestriero2026provingimprovingplatonicrepresentation}. By contrast, Ilya Sutskever and colleagues have argued that autoregressive prediction can foster broad, transferable abilities\cite{radford2019language,korbak2025cotmonitorability}. Crucially, resolving whether these abilities extend to quantitative physical reasoning has been hindered by a divide in evaluation: benchmarks designed for general-purpose language models typically rely on written question-answering\cite{rein2023gpqa}, code generation\cite{tian2024scicode}, abstract reasoning\cite{chollet2025arcagi2}, or tool use\cite{merrill2026terminalbench}, while spatial and physical tasks have largely been evaluated using specialized architectures \cite{kuznetsov2025nach0,zholus2025bindgpt}.

Molecular geometry provides a stringent setting to test spatial reasoning. A single molecule adopts different spatial shapes, known as conformers: its atoms and bonds remain identical, but their positions in space vary. As bonds rotate, atoms move closer together or farther apart, changing internal strain and enabling interactions such as intramolecular hydrogen bonds. The relative stability of these conformers determines their populations and influences the chemical, physical, and biological properties of a compound. Deciding which conformer is most stable requires integrating competing energetic contributions from its geometry. Models trained on structural data have demonstrated atomic-coordinate generation\cite{cavanagh2026structures,kuznetsov2025nach0,zholus2025bindgpt,flamshepherd2023three} and satisfaction of spatial constraints\cite{macdougall2026spatialconstraints}, while specialized machine-learned force fields can accurately evaluate conformational stability~\cite{batatia2023macehigherorderequivariant,10.1021/acs.jctc.0c00121}. Yet, how far general-purpose LLMs have acquired this ability without task-specific training remains unknown. More fundamentally, whether the scientific concepts encoded in human language impose systematic limitations also remains unexplored.

Here, we evaluate the ability of 28 LLMs to rank 810 conformers of 27 molecules with DFT reference energies. We test whether and how physical intuition has already emerged in frontier LLMs, and whether their predictions reflect the strengths and limitations of human-devised chemical concepts.

\begin{figure}[!htbp]
  \centering
  \includegraphics[width=\textwidth,height=0.62\textheight,keepaspectratio]{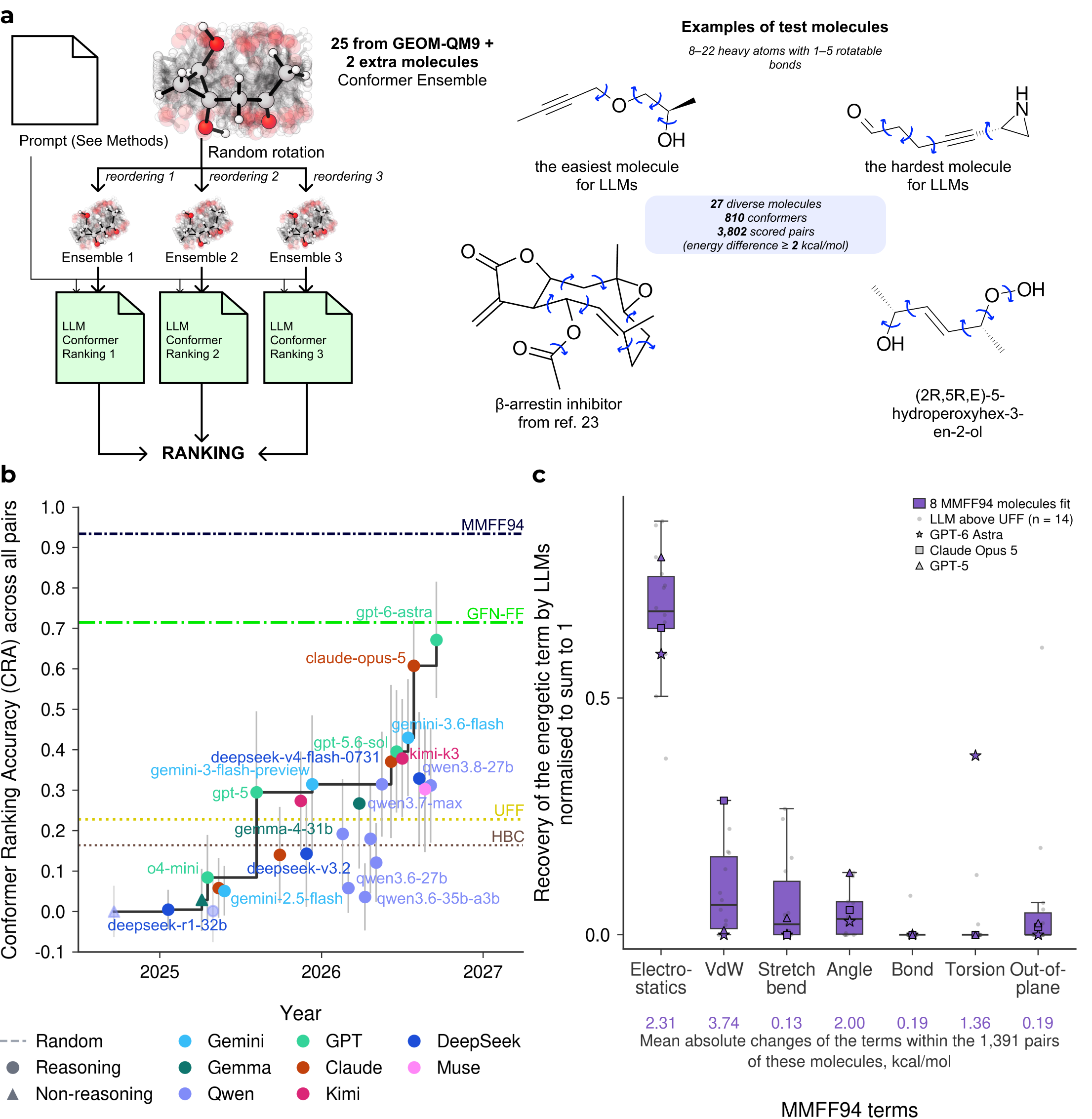}
  \caption{\textbf{Emergence of spatial physical reasoning in language models.} \textbf{a}, Geometry-only conformer ranking protocol and representative test molecules: easiest and hardest from GEOM-QM9 \cite{axelrod2022geom}, a \(\beta\)-arrestin inhibitor from Kahsai et al. \cite{kahsai2026small}, and hydroperoxide. Blue arrows denote bonds that rotate within the conformer ensemble. Each conformer ensemble was randomly rotated and ranked by LLMs in three independently shuffled runs against r\textsuperscript{2}SCAN-3c reference energies. \textbf{b}, Conformer Ranking Accuracy (CRA) across 3,802 reliable conformer pairs (with energy difference $\geq 2.0$ kcal/mol) versus model release date. Error bars denote molecule-level 95\% confidence intervals; colors identify model families; marker shapes distinguish reasoning configurations; horizontal lines indicate random, the hydrogen-bond counting rule (HBC; conformers with more intramolecular H-bonds ranked as more stable), and force-field baselines. The black line indicates the best CRA among all models released up to each date. Faded markers indicate Qwen 3 32B and Qwen 2.5 32B, evaluated on 26 and 25 molecules, respectively (see Supplementary Materials). \textbf{c}, Recovery of individual MMFF94 energy terms by LLMs across 1,391 pairs from 8 molecules with good agreement between MMFF94 and r\textsuperscript{2}SCAN-3c pairwise conformer energy differences. Boxplots show distributions across 14 LLMs with CRA above UFF in panel \textbf{b}.}
  \label{fig:benchmark}
\end{figure}

Each model was prompted to rank 30 conformers of the same molecule from atomic coordinates alone, without molecular identities, reference energies, or external tools (Fig.~\ref{fig:benchmark}a) and provide a textual explanation for its ranking. Reference rankings were obtained with r\textsuperscript{2}SCAN-3c \cite{Grimme2021_r2SCAN3c} and validated on a subset of molecules against DLPNO-CCSD(T)\cite{Riplinger2013} (see Methods). We quantified ranking agreement using Kendall’s \(\tau\), restricted to reliable conformer pairs (separated by at least 2.0~kcal~mol\ensuremath{{}^{-1}}), and refer to this metric as conformer ranking accuracy (CRA). A CRA of 1 indicates perfect agreement, whereas 0 corresponds to random ranking.

Most of the test molecules (25 out of 27) were extracted from the QM9 subset\cite{ramakrishnan2014qm9} of the GEOM\cite{axelrod2022geom} dataset. Because this dataset is publicly available and may have appeared in model training data, we assessed potential memorization by rotating conformer coordinates and separately testing whether models could recall the original GEOM energies. Their energy estimates were inaccurate and showed little relationship with ranking performance (Extended Data Fig.~\ref{fig:supplementary-energy-recall}). We additionally included two molecules whose conformers are absent from public data: a large \(\beta\)-arrestin inhibitor macrocycle from Kahsai et al. \cite{kahsai2026small} and (2R,5R,E)-5-hydroperoxyhex-3-en-2-ol. Both were selected for their conformational behavior: the macrocycle's stability is governed by an interplay of torsions that cannot all be satisfied simultaneously~\cite{Krivoshchapov2024}, whereas mono-peroxides lack the strong anomeric stabilization that organizes acetal conformations. Accordingly, rotation about the O--O bond has a relatively flat potential-energy profile: in dimethyl peroxide, a roughly $70^\circ$ range of C--O--O--C dihedrals lies within about 1~kcal~mol$^{-1}$ of the minimum \cite{gomes2015bisperoxides}. In both molecules, the conformational ordering is therefore set by a balance of competing contributions rather than by a single dominant interaction. All structures are shown in Extended Data Fig.~\ref{fig:supplementary-molecule-performance}a, with four examples in Fig.~\ref{fig:benchmark}a.

Strikingly, across the 28-model cohort, models released in 2024 and early 2025 generally ranked conformers close to random, whereas later releases produced increasingly accurate conformer rankings (Fig.~\ref{fig:benchmark}b). Several frontier models have matched or exceeded both the naive hydrogen-bond counting rule (HBC, CRA = 0.16), which simply ranks the conformer with more intramolecular hydrogen bonds as more stable, and the Universal Force Field (UFF, CRA = 0.23)\cite{Rappe1992}. And GPT-6 Astra\cite{openai2026gpt6astra} and Claude Opus 5\cite{anthropic2026claudeopus5} even reached CRA of 0.67 and 0.61, respectively, approaching the general-purpose, automatically parametrized GFN-FF force field (CRA = 0.72)\cite{Spicher2020_gfnff}. A physics-based MMFF94\cite{Halgren1996_MMFF94} force field was substantially more accurate on the selected relatively simple organic molecules with CRA of 0.93.

However, the aggregate CRA does not tell why LLMs succeed or fail. To identify which physical interactions each LLM gets right, we selected the eight molecules for which total MMFF94 energy differences reproduced the r\textsuperscript{2}SCAN-3c pairwise energy differences with $R^2\ge0.8$ (see Extended Data Fig.~\ref{fig:supplementary-mmff-alignment}). For each LLM separately, we fitted the probability that it ranks a conformer pair correctly, $P(\mathrm{correct})=\sigma\!\left(b_m+\sum_t c_t\Delta E_t\right)$, where $\sigma$ is the logistic function and $\Delta E_t$ is the contribution of MMFF94 term $t$ to the pair’s energy difference, with one intercept per molecule and non-negative coefficients for the seven MMFF94 terms (see the x-axis in Fig.~\ref{fig:benchmark}c). Because some terms varied little within pairs (e.g., stretch--bend, bond, and out-of-plane; see values below the x-axis in Fig.~\ref{fig:benchmark}c), their coefficients were weakly constrained by the data, so we therefore used L2 regularization ($\lambda=0.01$). The fitted coefficients were normalized to sum to one; they indicate each term's share in the rankings (Fig.~\ref{fig:benchmark}c).

Figure~\ref{fig:benchmark}c shows the recovery-factor distributions for the 14 LLMs that outperform UFF in Fig.~\ref{fig:benchmark}b. Most of these LLMs recover large portions of electrostatics, which is energetically important (2.31 kcal/mol average absolute variation within pairs) and convenient: it can be inferred from the positions of only two atoms and is well conceptualized and quantified in chemical literature. The other energy terms are virtually omitted by most LLMs. Interestingly, Claude Opus 5 and GPT-6 Astra, the top overall performers (Fig.~\ref{fig:benchmark}b), but each recovers a different extra term: Opus 5 recovers \textasciitilde{}30\% of van der Waals (VdW) interactions, which have the largest average absolute variation (3.74 kcal/mol), whereas Astra omits VdW completely but recovers \textasciitilde{}40\% of torsion strain—roughly three times the next-highest recovery (0.13 for Gemini 3 Flash). Thus, even frontier LLMs have not acquired accurate internal representations of chemical interactions within molecules; instead, they have learned some key concepts that human chemists use to understand molecules, which already allows them to reach the accuracy of the GFN-FF force field.

\begin{figure}[H]
  \centering
  \includegraphics[width=\textwidth,height=0.72\textheight,keepaspectratio]{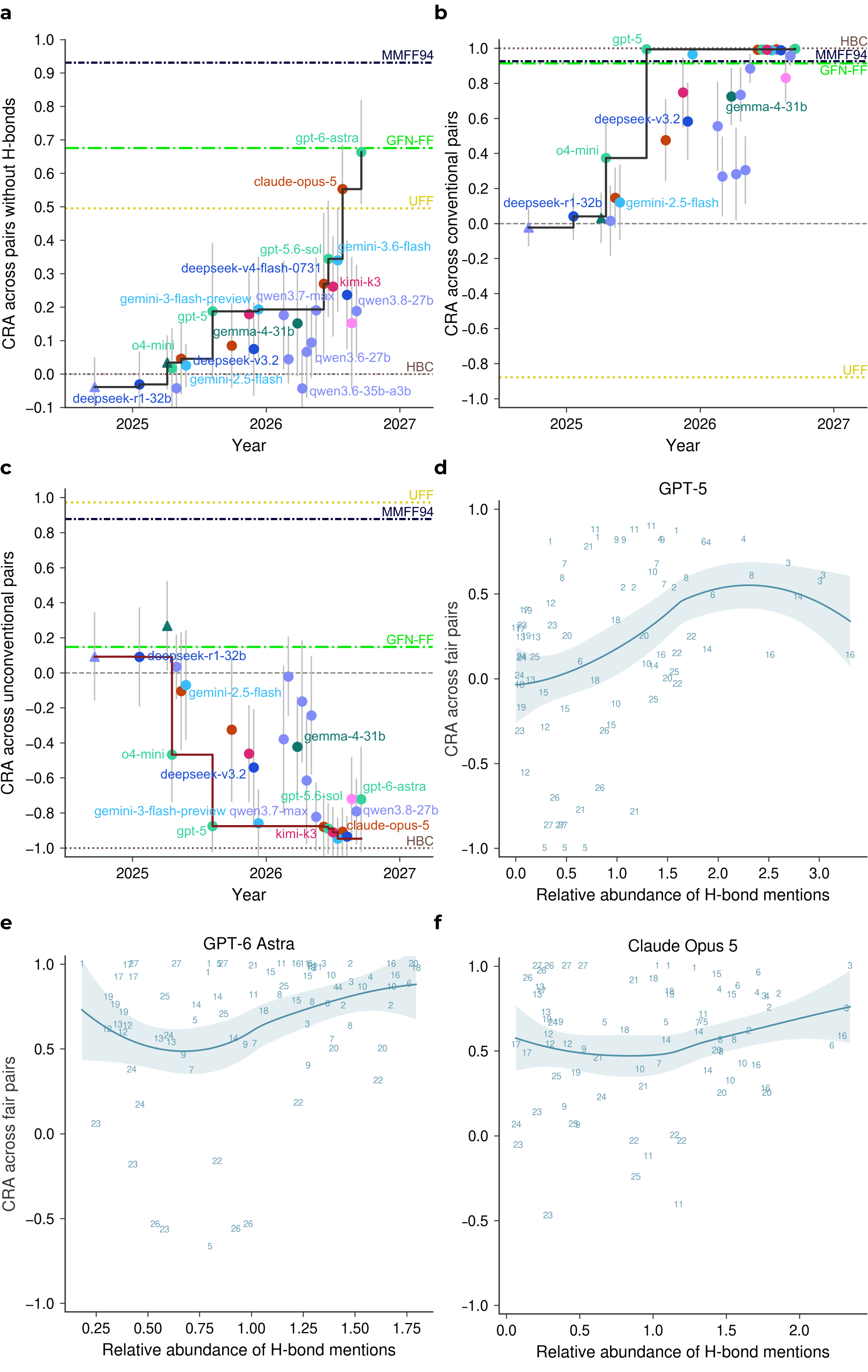}
  \caption{\textbf{Hydrogen-bond bias and its relationship to physical reasoning accuracy}. \textbf{a–c}, Conformer Ranking Accuracy (CRA) versus model release date on pairs without hydrogen bonds in either conformer (no H-bond; 2,739 pairs; \textbf{a}), pairs where the more hydrogen-bonded conformer is more stable (conventional; 715 pairs; \textbf{b}), and pairs where it is less stable (unconventional; 90 pairs; \textbf{c}). Error bars denote molecule-level 95\% confidence intervals; horizontal lines indicate random, the hydrogen-bond counting rule (HBC; conformers with more intramolecular H-bonds ranked as more stable), and force-field baselines. Black staircase indicates the best CRA among models released up to each date (\textbf{a}, \textbf{b}); the red staircase indicates the lowest CRA among models released up to each date (\textbf{c}). Model families and reasoning configurations are identified as in Fig.~\ref{fig:benchmark}b. \textbf{d–f}, CRA on pairs with equal hydrogen-bond counts (fair; 2,997 pairs) versus the relative abundance of hydrogen-bond mentions in model explanations for GPT-5 (\textbf{d}), GPT-6 Astra (\textbf{e}), and Claude Opus 5 (\textbf{f}). Numbers identify molecules in individual runs; curves show LOESS fits with 95\% confidence intervals. Metric definitions and filtering details are provided in Methods.}
  \label{fig:ranking-heatmap}
\end{figure}

Good recovery of electrostatics by LLMs led us to examine its influence on their conformer ranking performance. Of the 3,802 reliable conformer pairs, 2,739 have no intramolecular hydrogen bonds in either conformer. On these pairs, the CRA of frontier models still increases with release date (Fig.~\ref{fig:ranking-heatmap}a), demonstrating that successful ranking extends beyond identifying hydrogen bonds. For 805 pairs with differing hydrogen-bond counts, however, a simple chemical heuristic becomes highly informative. In 715 such pairs (89\%), the conformer with more hydrogen bonds is more stable. We designate these pairs as conventional: models can rank them correctly by recognizing and counting hydrogen bonds, without evaluating the full energetic balance. Here, many recent LLMs achieve nearly perfect CRA (Fig.~\ref{fig:ranking-heatmap}b). A single intramolecular hydrogen bond can stabilize a conformer by a few kcal mol\ensuremath{{}^{-1}} \cite{steiner2002hydrogenbond}, making this heuristic physically plausible in many cases.

The same heuristic, however, fails for the remaining 90 pairs (11\%), in which the more hydrogen-bonded conformer is less stable. These unconventional pairs expose a systematic reversal: models that rank conventional pairs more accurately tend to perform worse when the heuristic fails (Fig.~\ref{fig:ranking-heatmap}c). Notably, GPT-6 Astra is a partial exception, retaining a CRA of -0.72 on unconventional pairs while performing strongly on conventional pairs – this behavior is likely linked to its ability to partially recover distributed effects (such as torsional strain, angle bending, etc.), which change substantially in most of the unconventional pairs (Extended Data Fig.~\ref{fig:supplementary-distributed-strain}). Nevertheless, its negative CRA shows that the hydrogen-bond heuristic persists even in frontier models. The GFN-FF force field also performs close to random on the unconventional pairs, showing that these cases challenge machine learning force fields as well.

\clearpage

To examine performance beyond differences in hydrogen-bond counts, we considered the 2,997 pairs with equal counts in both conformers (fair pairs). Their ranking accuracy varies substantially across molecules (Extended Data Fig.~\ref{fig:supplementary-molecule-performance}b). We therefore analyzed the textual explanations accompanying model rankings and compared CRA on fair pairs with the relative frequency of hydrogen-bond mentions for each model (Fig.~\ref{fig:ranking-heatmap}d–f). GPT-5 only shows accurate rankings for molecules where it frequently mentions hydrogen bonds in its explanations; GPT-6 Astra and Claude Opus 5 sometimes work well for molecules even lacking (recognized) hydrogen bonds but achieve stable high accuracy only when hydrogen bonds are mentioned frequently. Across models, chemical vocabulary is informative but not sufficient: frequent mentions of hydrogen bonds do not uniformly imply high accuracy for weaker models (Extended Data Fig.~\ref{fig:supplementary-hbond-models}).

\begin{figure}[!htbp]
  \centering
  \includegraphics[width=\textwidth,height=0.68\textheight,keepaspectratio]{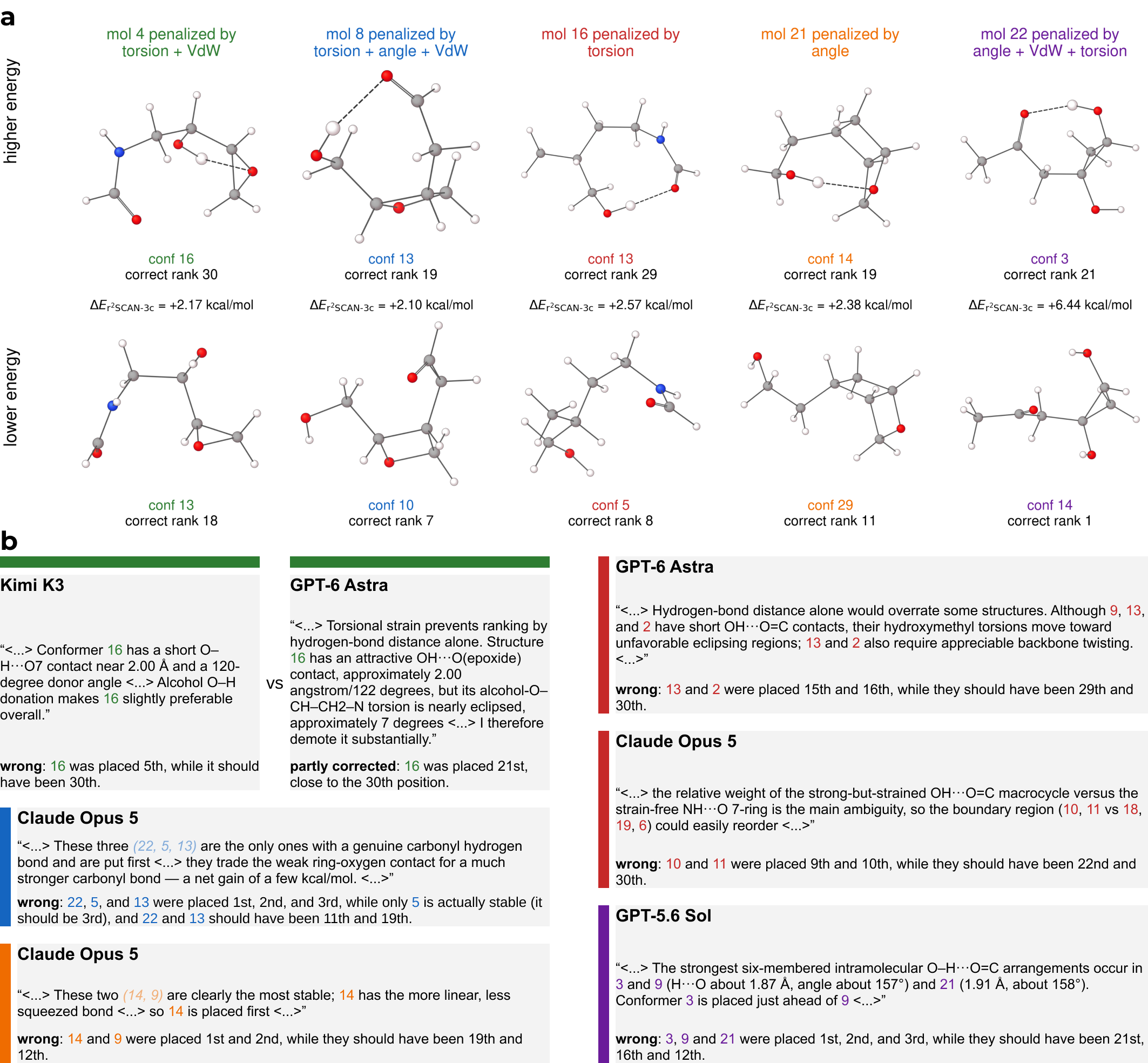}
  \caption{\textbf{Recognition of chemical interactions does not guarantee accurate energetic predictions}. \textbf{a}, Representative unconventional conformer pairs with hydrogen bonds outweighed by torsional, angle or van der Waals penalties, identified from MMFF94 decomposition. Relative r\textsuperscript{2}SCAN-3c energies are indicated; dashed lines denote hydrogen bonds. \textbf{b}, Excerpts from LLM explanations illustrating correct and incorrect energetic assessments despite recognition of chemical interactions.}
  \label{fig:hbond-mentions}
\end{figure}
\clearpage

In unconventional pairs, the local stabilization associated with additional hydrogen bonds is outweighed by other energetic contributions. Representative examples reveal torsional, angle-bending and van der Waals penalties that offset electrostatically favored contacts (Fig.~\ref{fig:hbond-mentions}a). Model explanations often recognize these interactions and quantify their geometry, yet differ in the energetic significance assigned to them (Fig.~\ref{fig:hbond-mentions}b). In one example, Kimi K3 favors the hydrogen-bonded conformer (Fig.~\ref{fig:hbond-mentions}b, left green) while GPT-6 Astra identifies a competing torsional penalty, which it is good at (Fig.~\ref{fig:benchmark}c) and demotes it (Fig.~\ref{fig:hbond-mentions}b, right green).

A complementary, hydrogen-bond-free example is the hydroperoxide. Although it carries both OH and OOH groups, the E-configured alkene holds them apart, and none of its 30 conformers forms an intramolecular hydrogen bond. Across its 104 reliable pairs, four MMFF94 terms are comparable in magnitude (torsion 1.54, angle 1.45, van der Waals 1.09, and electrostatic 0.97~kcal~mol$^{-1}$), so no single term determines the ordering. Taken together, these terms nevertheless yield strong agreement with the reference: MMFF94 reaches CRA = 0.67 and UFF 0.62, the latter far above its benchmark average of 0.23. Among LLMs, GPT-6 Astra performs best (CRA = 0.62), consistent with its being the only model to recover a substantial fraction of the torsional term (Fig.~\ref{fig:benchmark}c). GFN-FF, by contrast, anti-correlates with the reference ordering on this molecule (CRA = -0.27, against 0.72 across the benchmark).

Thus, frontier general-purpose LLMs can infer aspects of molecular stability directly from three-dimensional atomic coordinates. Yet instead of acquiring quantitative understanding of intramolecular interactions like those built into physics-based force fields, they have learned the easily quantifiable chemical concepts, just like humans do. Thus, autoregressive prediction of human-written texts did not so far result in accurate molecular-physics world models in frontier LLMs, consistent with the world-model critique\cite{lecun2022path,balestriero2026provingimprovingplatonicrepresentation}. The models therefore appear to inherit a simplified, human-like physical prior, shaped by literature that emphasizes electrostatics and hydrogen bonding far more than distributed strain. On the other hand, the example of GPT-6 Astra shows that LLMs are in principle able to learn how to quantify distributed interactions, such as torsion strain. While LLMs still struggle with such interactions, physics-informed models offer a complementary route to quantitative prediction where human-derived verbal concepts are insufficient, highlighting their value as tools for scientific LLMs.\cite{medvedev2026chemistryai}

Our findings open new opportunities for language models in molecular design and autonomous AI-driven chemistry \cite{coley2020autonomous,tom2024selfdriving,ding2026survey}, while also exposing limitations arising from their reliance on human scientific language. Since the human approach to developing and expressing scientific concepts is largely shared across disciplines, these findings are likely to extend beyond chemistry.

\FloatBarrier

\bibliographystyle{unsrturl}
\bibliography{references}

@misc{lecun2022path,
  author       = {Y. LeCun},
  title        = {A Path Towards Autonomous Machine Intelligence},
  year         = {2022},
  url          = {https://openreview.net/forum?id=BZ5a1r-kVsf}
}

@misc{klindt2026lejepa,
  author       = {D. Klindt and Y. LeCun and R. Balestriero},
  title        = {When Does {LeJEPA} Learn a World Model?},
  year         = {2026},
  eprint       = {2605.26379},
  archivePrefix = {arXiv},
  doi          = {10.48550/arXiv.2605.26379}
}

@misc{balestriero2026provingimprovingplatonicrepresentation,
  author       = {Randall Balestriero and Yann LeCun},
  title        = {Proving and Improving the Platonic Representation Hypothesis with {JEPAs}},
  year         = {2026},
  eprint       = {2610.39302},
  archivePrefix = {arXiv},
  primaryClass = {cs.AI},
  url          = {https://arxiv.org/abs/2610.39302}
}

@misc{radford2019language,
  author       = {A. Radford and J. Wu and R. Child and D. Luan and D. Amodei and I. Sutskever},
  title        = {Language Models Are Unsupervised Multitask Learners},
  year         = {2019},
  publisher     = {OpenAI}
}

@misc{korbak2025cotmonitorability,
  author       = {T. Korbak and M. Balesni and E. Barnes and Y. Bengio and others},
  title        = {Chain of Thought Monitorability: A New and Fragile Opportunity for {AI} Safety},
  year         = {2025},
  eprint       = {2507.11473},
  archivePrefix = {arXiv},
  doi          = {10.48550/arXiv.2507.11473}
}

@misc{rein2023gpqa,
  author       = {D. Rein and B. L. Hou and A. C. Stickland and J. Petty and R. Y. Pang and J. Dirani and J. Michael and S. R. Bowman},
  title        = {{GPQA}: A Graduate-Level Google-Proof Q\&A Benchmark},
  year         = {2023},
  eprint       = {2311.12022},
  archivePrefix = {arXiv},
  doi          = {10.48550/arXiv.2311.12022}
}

@misc{tian2024scicode,
  author       = {M. Tian and L. Gao and S. D. Zhang and X. Chen and C. Fan and X. Guo and R. Haas and P. Ji and K. Krongchon and Y. Li and others},
  title        = {{SciCode}: A Research Coding Benchmark Curated by Scientists},
  year         = {2024},
  eprint       = {2407.13168},
  archivePrefix = {arXiv},
  doi          = {10.48550/arXiv.2407.13168}
}

@misc{chollet2025arcagi2,
  author       = {F. Chollet and M. Knoop and G. Kamradt and B. Landers and H. Pinkard},
  title        = {{ARC-AGI-2}: A New Challenge for Frontier {AI} Reasoning Systems},
  year         = {2025},
  eprint       = {2505.11831},
  archivePrefix = {arXiv},
  doi          = {10.48550/arXiv.2505.11831}
}

@inproceedings{merrill2026terminalbench,
  author       = {M. A. Merrill and A. G. Shaw and N. Carlini and B. Li and H. Raj and I. Bercovich and L. Shi and J. Y. Shin and T. Walshe and E. K. Buchanan and J. Shen and G. Ye and H. Lin and J. Poulos and M. Wang and M. Nezhurina and D. Lu and O. M. Mastromichalakis and Z. Xu and Z. Chen and Y. Liu and R. Zhang and L. L. Chen and A. Kashyap and J.-L. Uslu and J. Li and J. Wu and M. Yan and S. Bian and V. Sharma and K. Sun and S. Dillmann and A. Anand and A. Lanpouthakoun and B. Koopah and C. Hu and E. K. Guha and G. H. S. Dreiman and J. Zhu and K. Krauth and L. Zhong and N. Muennighoff and R. K. Amanfu and S. Tan and S. Pimpalgaonkar and T. Aggarwal and X. Lin and X. Lan and X. Zhao and Y. Liang and Y. Wang and Z. Wang and C. Zhou and D. Heineman and H. Liu and H. Trivedi and J. Yang and J. Lin and M. Shetty and M. Yang and N. Omi and N. Raoof and S. Li and T. Y. Zhuo and W. Lin and Y. Dai and Y. Wang and W. Chai and S. Zhou and D. Wahdany and Z. She and J. Hu and Z. Dong and Y. Zhu and S. Cui and A. Saiyed and A. Kolbeinsson and C. M. Rytting and R. Marten and Y. Wang and J. Jitsev and A. Dimakis and A. Konwinski and L. Schmidt},
  title        = {{Terminal-bench}: Benchmarking Agents on Hard, Realistic Tasks in Command Line Interfaces},
  booktitle    = {The Fourteenth International Conference on Learning Representations},
  year         = {2026},
  url          = {https://openreview.net/forum?id=a7Qa4CcHak}
}

@article{ramakrishnan2014qm9,
  author       = {R. Ramakrishnan and P. O. Dral and M. Rupp and O. A. von Lilienfeld},
  title        = {Quantum Chemistry Structures and Properties of 134 Kilo Molecules},
  journal      = {Scientific Data},
  volume       = {1},
  pages        = {140022},
  year         = {2014}
}

@article{axelrod2022geom,
  author       = {S. Axelrod and R. G{\'o}mez-Bombarelli},
  title        = {{GEOM}, Energy-Annotated Molecular Conformations for Property Prediction and Molecular Generation},
  journal      = {Scientific Data},
  volume       = {9},
  pages        = {185},
  year         = {2022}
}

@misc{cavanagh2026structures,
  author       = {J. M. Cavanagh and J. B. Arnold and G. B. Alteri and A. Gritsevskiy and T. Head-Gordon},
  title        = {How Well Can Frontier Large Language Models Generate Structures? High Quality Prediction of Molecular Geometries with Help from Fine-Tuning},
  year         = {2026},
  eprint       = {2607.13350},
  archivePrefix = {arXiv},
  doi          = {10.48550/arXiv.2607.13350}
}

@inproceedings{kuznetsov2025nach0,
  author       = {M. Kuznetsov and A. Valiev and A. Aliper and D. Polykovskiy and E. Tutubalina and R. Shayakhmetov and Z. Miftahutdinov},
  title        = {{nach0-pc}: Multi-Task Language Model with Molecular Point Cloud Encoder},
  booktitle    = {Proceedings of the AAAI Conference on Artificial Intelligence},
  volume       = {39},
  pages        = {24357--24365},
  year         = {2025}
}

@inproceedings{zholus2025bindgpt,
  author       = {A. Zholus and M. Kuznetsov and R. Schutski and R. Shayakhmetov and D. Polykovskiy and S. Chandar and A. Zhavoronkov},
  title        = {{BindGPT}: A Scalable Framework for 3D Molecular Design via Language Modeling and Reinforcement Learning},
  booktitle    = {Proceedings of the AAAI Conference on Artificial Intelligence},
  volume       = {39},
  pages        = {26083--26091},
  year         = {2025}
}

@misc{flamshepherd2023three,
  author       = {D. Flam-Shepherd and A. Aspuru-Guzik},
  title        = {Language Models Can Generate Molecules, Materials, and Protein Binding Sites Directly in Three Dimensions as {XYZ}, {CIF}, and {PDB} Files},
  year         = {2023},
  eprint       = {2305.05708},
  archivePrefix = {arXiv},
  doi          = {10.48550/arXiv.2305.05708}
}

@misc{macdougall2026spatialconstraints,
  author       = {T. MacDougall and M. Kuznetsov and R. Schutski and R. Shayakhmetov and M. Malkov and V. Aladinskiy and A. Aliper and A. Zhavoronkov},
  title        = {Do Language Models Dream of Binding Molecules? Benchmarking {LLMs} under Spatial Constraints},
  year         = {2026},
  eprint       = {2607.18144},
  archivePrefix = {arXiv},
  doi          = {10.48550/arXiv.2607.18144}
}

@article{kahsai2026small,
  author       = {A. W. Kahsai and N. Pakharukova and H. Y. Kwon and K. S. Shah and others},
  title        = {Small-Molecule Modulation of $\beta$-Arrestins},
  journal      = {Nature},
  volume       = {656},
  pages        = {770--779},
  year         = {2026}
}

@article{Grimme2021_r2SCAN3c,
  author       = {S. Grimme and A. Hansen and S. Ehlert and J.-M. Mewes},
  title        = {$r^2${SCAN-3c}: A ``Swiss Army Knife'' Composite Electronic-Structure Method},
  journal      = {The Journal of Chemical Physics},
  volume       = {154},
  pages        = {064103},
  year         = {2021}
}

@article{Riplinger2013,
  author       = {C. Riplinger and B. Sandhoefer and A. Hansen and F. Neese},
  title        = {Natural Triple Excitations in Local Coupled Cluster Calculations with Pair Natural Orbitals},
  journal      = {The Journal of Chemical Physics},
  volume       = {139},
  pages        = {134101},
  year         = {2013}
}

@article{Rappe1992,
  author       = {A. K. Rapp{\'e} and C. J. Casewit and K. S. Colwell and W. A. Goddard III and W. M. Skiff},
  title        = {{UFF}, a Full Periodic Table Force Field for Molecular Mechanics and Molecular Dynamics Simulations},
  journal      = {Journal of the American Chemical Society},
  volume       = {114},
  pages        = {10024--10035},
  year         = {1992}
}

@misc{openai2026gpt6astra,
  author       = {{OpenAI}},
  title        = {{GPT-6 Astra}: A New Generation of Intelligence},
  year         = {2026},
  url          = {https://openai.com/index/gpt-6-astra/}
}

@misc{anthropic2026claudeopus5,
  author       = {{Anthropic}},
  title        = {Introducing Claude Opus 5},
  year         = {2026},
  url          = {https://www.anthropic.com/news/claude-opus-5}
}

@article{Spicher2020_gfnff,
  author       = {S. Spicher and S. Grimme},
  title        = {Robust Atomistic Modeling of Materials, Organometallic, and Biochemical Systems},
  journal      = {Angewandte Chemie International Edition},
  volume       = {59},
  pages        = {15665--15673},
  year         = {2020}
}

@article{Halgren1996_MMFF94,
  author       = {T. A. Halgren},
  title        = {Merck Molecular Force Field. {I}. Basis, Form, Scope, Parameterization, and Performance of {MMFF94}},
  journal      = {Journal of Computational Chemistry},
  volume       = {17},
  pages        = {490--519},
  year         = {1996}
}

@misc{batatia2023macehigherorderequivariant,
  title        = {{MACE}: Higher Order Equivariant Message Passing Neural Networks for Fast and Accurate Force Fields},
  author       = {Ilyes Batatia and D{\'a}vid P{\'e}ter Kov{\'a}cs and Gregor N. C. Simm and Christoph Ortner and G{\'a}bor Cs{\'a}nyi},
  year         = {2023},
  eprint       = {2206.07697},
  archivePrefix = {arXiv},
  primaryClass = {stat.ML},
  url          = {https://arxiv.org/abs/2206.07697}
}

@article{10.1021/acs.jctc.0c00121,
  author       = {Christian Devereux and Justin S. Smith and Kate K. Huddleston and Kipton Barros and Roman Zubatyuk and Olexandr Isayev and Adrian E. Roitberg},
  title        = {Extending the Applicability of the {ANI} Deep Learning Molecular Potential to Sulfur and Halogens},
  journal      = {Journal of Chemical Theory and Computation},
  volume       = {16},
  number       = {7},
  pages        = {4192--4202},
  year         = {2020},
  doi          = {10.1021/acs.jctc.0c00121},
  url          = {https://doi.org/10.1021/acs.jctc.0c00121}
}

@article{steiner2002hydrogenbond,
  author       = {T. Steiner},
  title        = {The Hydrogen Bond in the Solid State},
  journal      = {Angewandte Chemie International Edition},
  volume       = {41},
  pages        = {48--76},
  year         = {2002},
  doi          = {10.1002/1521-3773(20020104)41:1<48::AID-ANIE48>3.0.CO;2-U}
}

@article{medvedev2026chemistryai,
  author       = {Medvedev, Michael G. and Serov, Nikita S. and Chaliy, Vasily A. and Ilin, Egor A. and Prolomov, Ilya V. and Eremeyeva, Maria A. and Schneider, Mark Yu. and Mezentsev, Igor A. and Artamonov, Daniil V. and Boiko, Daniil A. and Vanovskiy, Vladimir V. and Ryabov, Alexander A. and Burnaev, Evgeny V. and Ananikov, Valentine P.},
  title        = {Chemistry-transforming applications of artificial intelligence},
  journal      = {Chemical Society Reviews},
  year         = {2026},
  month        = {nov},
  issn         = {0306-0012},
  doi          = {10.1039/d4cs01100g},
  url          = {https://doi.org/10.1039/d4cs01100g}
}

@article{klimesch2026designingphysics,
  author       = {Klimesch, J. and Arlt, S. and Ruiz-Gonzalez, C. and others},
  title        = {Designing physics experiments with artificial intelligence},
  journal      = {Nature},
  volume       = {657},
  pages        = {47--58},
  year         = {2026},
  month        = {sep},
  doi          = {10.1038/s41586-026-10898-6},
  url          = {https://doi.org/10.1038/s41586-026-10898-6}
}

@article{dip2026llmagents,
  author       = {Dip, Sajib Acharjee and Mallick, Dipanwita and Acharjee Shuvo, Uddip and Barua Soumma, Shovito and Rafsani, Fazle and Kumar Paul, Bikash and Ahmed Moumi, Nazifa and Ahmed, Shafayat and Zhang, Liqing},
  title        = {Large language model agents for biological intelligence across genomics, proteomics, spatial biology, and biomedicine},
  journal      = {Briefings in Bioinformatics},
  volume       = {27},
  number       = {2},
  pages        = {bbag110},
  year         = {2026},
  month        = {mar},
  issn         = {1477-4054},
  doi          = {10.1093/bib/bbag110},
  url          = {https://doi.org/10.1093/bib/bbag110}
}

@misc{rdkit,
  author       = {Greg Landrum},
  title        = {{RDKit}: Open-Source Cheminformatics, Version 2024.03.3},
  year         = {2024},
  doi          = {10.5281/zenodo.11396708},
  url          = {https://doi.org/10.5281/zenodo.11396708}
}

@article{kay1984sapirwhorf,
  author       = {P. Kay and W. Kempton},
  title        = {What Is the Sapir--Whorf Hypothesis?},
  journal      = {American Anthropologist},
  volume       = {86},
  pages        = {65--79},
  year         = {1984}
}

@article{coley2020autonomous,
  author       = {C. W. Coley and N. S. Eyke and K. F. Jensen},
  title        = {Autonomous Discovery in the Chemical Sciences Part {I}: Progress},
  journal      = {Angewandte Chemie International Edition},
  volume       = {59},
  pages        = {22858--22893},
  year         = {2020}
}

@article{tom2024selfdriving,
  author       = {G. Tom and S. P. Schmid and S. G. Baird and Y. Cao and K. Darvish and H. Hao and S. Lo and S. Pablo-Garc{\'i}a and E. M. Rajaonson and M. Skreta and others},
  title        = {Self-Driving Laboratories for Chemistry and Materials Science},
  journal      = {Chemical Reviews},
  volume       = {124},
  pages        = {9633--9732},
  year         = {2024}
}

@article{ding2026survey,
  author       = {H. Ding and P. Hua and Z. Huang},
  title        = {Survey on Recent Progress of {AI} for Chemistry: Methods, Applications, and Opportunities},
  journal      = {Frontiers of Computer Science},
  volume       = {20},
  pages        = {2011358},
  year         = {2026}
}

@article{Wang2020_ETKDG,
  author       = {S. Wang and J. Witek and G. A. Landrum and S. Riniker},
  title        = {Improving Conformer Generation for Small Rings and Macrocycles Based on Distance Geometry and Experimental Torsional-Angle Preferences},
  journal      = {Journal of Chemical Information and Modeling},
  volume       = {60},
  pages        = {2044--2058},
  year         = {2020}
}

@article{Pracht2024_CREST,
  author       = {P. Pracht and S. Grimme and C. Bannwarth and F. Bohle and S. Ehlert and G. Feldmann and J. Gorges and M. M{\"u}ller and T. Neudecker and C. Plett and S. Spicher and P. Steinbach and P. A. Weso{\l}owski and F. Zeller},
  title        = {{CREST}---A Program for the Exploration of Low-Energy Molecular Chemical Space},
  journal      = {The Journal of Chemical Physics},
  volume       = {160},
  pages        = {114110},
  year         = {2024}
}

@article{Neese2025orca6,
  author       = {Frank Neese},
  title        = {Software Update: The {ORCA} Program System---Version 6.0},
  journal      = {WIREs Computational Molecular Science},
  volume       = {15},
  pages        = {e70019},
  year         = {2025}
}

@article{Weininger1988smiles,
  author       = {D. Weininger},
  title        = {{SMILES}, a Chemical Language and Information System. 1. Introduction to Methodology and Encoding Rules},
  journal      = {Journal of Chemical Information and Computer Sciences},
  volume       = {28},
  pages        = {31--36},
  year         = {1988}
}

@article{xtb,
  author       = {C. Bannwarth and E. Caldeweyher and S. Ehlert and A. Hansen and P. Pracht and J. Seibert and S. Spicher and S. Grimme},
  title        = {Extended Tight-Binding Quantum Chemistry Methods},
  journal      = {WIREs Computational Molecular Science},
  volume       = {11},
  pages        = {e1493},
  year         = {2021},
  doi          = {10.1002/wcms.1493}
}

@article{Bannwarth2019,
  author       = {C. Bannwarth and S. Ehlert and S. Grimme},
  title        = {{GFN2-xTB}---An Accurate and Broadly Parametrized Self-Consistent Tight-Binding Quantum Chemical Method with Multipole Electrostatics and Density-Dependent Dispersion Contributions},
  journal      = {Journal of Chemical Theory and Computation},
  volume       = {15},
  pages        = {1652--1671},
  year         = {2019}
}

@article{liu-etal-2024-lost,
  author       = {N. F. Liu and K. Lin and J. Hewitt and A. Paranjape and M. Bevilacqua and F. Petroni and P. Liang},
  title        = {Lost in the Middle: How Language Models Use Long Contexts},
  journal      = {Transactions of the Association for Computational Linguistics},
  volume       = {12},
  pages        = {157--173},
  year         = {2024}
}

@misc{pipis2025loop,
  author       = {C. Pipis and S. Garg and V. Kontonis and V. Shrivastava and A. Krishnamurthy and D. Papailiopoulos},
  title        = {Wait, Wait, Wait... Why Do Reasoning Models Loop?},
  year         = {2025},
  eprint       = {2512.12895},
  archivePrefix = {arXiv},
  doi          = {10.48550/arXiv.2512.12895}
}

@article{Weigend2005,
  author       = {F. Weigend and R. Ahlrichs},
  title        = {Balanced Basis Sets of Split Valence, Triple Zeta Valence and Quadruple Zeta Valence Quality for {H} to {Rn}: Design and Assessment of Accuracy},
  journal      = {Physical Chemistry Chemical Physics},
  volume       = {7},
  pages        = {3297--3305},
  year         = {2005}
}

@article{Liakos2015,
  author       = {D. G. Liakos and M. Sparta and M. K. Kesharwani and J. M. L. Martin and F. Neese},
  title        = {Exploring the Accuracy Limits of Local Pair Natural Orbital Coupled-Cluster Theory},
  journal      = {Journal of Chemical Theory and Computation},
  volume       = {11},
  pages        = {1525--1539},
  year         = {2015}
}

@article{Neese2009rijcosx,
  author       = {F. Neese and F. Wennmohs and A. Hansen and U. Becker},
  title        = {Efficient, Approximate and Parallel Hartree--Fock and Hybrid {DFT} Calculations. A ``Chain-of-Spheres'' Algorithm for the Hartree--Fock Exchange},
  journal      = {Chemical Physics},
  volume       = {356},
  pages        = {98--109},
  year         = {2009}
}

@article{Hattig2005,
  author       = {C. H{\"a}ttig},
  title        = {Optimization of Auxiliary Basis Sets for {RI-MP2} and {RI-CC2} Calculations: Core--Valence and Quintuple-$\zeta$ Basis Sets for {H} to {Ar} and {QZVPP} Basis Sets for {Li} to {Kr}},
  journal      = {Physical Chemistry Chemical Physics},
  volume       = {7},
  pages        = {59--66},
  year         = {2005}
}

@article{gomes2015bisperoxides,
  author       = {Gomes, Gabriel dos Passos and Vil', Vera and Terent'ev, Alexander and Alabugin, Igor V.},
  title        = {Stereoelectronic source of the anomalous stability of bis-peroxides},
  journal      = {Chemical Science},
  volume       = {6},
  number       = {12},
  pages        = {6783--6791},
  year         = {2015},
  month        = {dec},
  doi          = {10.1039/c5sc02402a},
  url          = {https://doi.org/10.1039/c5sc02402a}
}

@article{Krivoshchapov2024,
  author       = {Krivoshchapov, Nikolai V. and Medvedev, Michael G.},
  title        = {Ring kinematics-informed conformation space exploration},
  journal      = {WIREs Computational Molecular Science},
  volume       = {14},
  number       = {1},
  pages        = {e1690},
  year         = {2024},
  doi          = {10.1002/wcms.1690},
  url          = {https://doi.org/10.1002/wcms.1690}
}

\section*{Methods}

\subsection*{Extraction of data from GEOM}

Conformer ensembles were taken from the QM9 subset of GEOM~\cite{axelrod2022geom}. The molecular records were shuffled, and the first 30 molecules containing at least 30 unique conformers were selected. For each molecule, the 30 lowest-energy conformers reported by GEOM were retained. Each of the resulting 900 geometries was then rigidly rotated by an independent matrix $R = R_z R_y R_x$, with Euler angles drawn from $\mathrm{Uniform}(0,2\pi)$. All selection and rotation parameters, random states, and software dependencies are fixed in the accompanying reproducible environment, which reconstructs the dataset exactly (see the Data Availability section). No atoms were added, removed, or permuted, and the atom order was held identical across all 30 conformers within each ensemble.

After preparing the ensembles, single-point r\textsuperscript{2}SCAN-3c energies were obtained for all $30\times30=900$ geometries. Before defining the final scored cohort, we ordered the 30 molecules by increasing r\textsuperscript{2}SCAN-3c energy span, defined as the difference between the highest- and lowest-energy conformers in each ensemble, and selected ten molecules approximately evenly spaced across this ordering (with a small random jitter) for DLPNO-CCSD(T) validation. The resulting ten-molecule subset therefore covered almost the entire range from narrow to broad conformational energy landscapes.

The observed agreement between r\textsuperscript{2}SCAN-3c and DLPNO-CCSD(T) conformer-pair orderings was perfect at the 1.5 and 2.0~kcal~mol$^{-1}$ thresholds (100.0\% and 100.0\%, respectively) and remained near-perfect at 1.0~kcal~mol$^{-1}$ (99.9\%). We then excluded five molecules with fewer than 15 conformer pairs satisfying $|\Delta E|\ge2.0$~kcal~mol$^{-1}$. Requiring at least 15 energetically well-separated pairs ensured that each molecule contributed enough unambiguous comparisons for an informative molecule-level ranking score, preventing $\tau$ from being determined by only a few pairs. The filtering criterion depended only on the r\textsuperscript{2}SCAN-3c energy gaps and was independent of both the DLPNO-CCSD(T) agreement and the model rankings. The GEOM-QM9 part of LLMConfBench contained 25 molecules, although prompts were generated for all 30 ensembles.

\subsection*{Conformer generation of two extra molecules}

We also augmented the dataset with a large $\beta$-arrestin inhibitor macrocycle from ref.~\cite{kahsai2026small} and $(2R,5R,E)$-5-hydroperoxyhex-3-en-2-ol (both absent from QM9 and CAS SciFinder) to test how models perform on out-of-distribution molecules. Thus, the final dataset contained 27 molecules and 810 geometries (Supplementary Table~\ref{tab:supplementary-descriptors} and Extended Data Figs.~\ref{fig:supplementary-molecule-performance}, \ref{fig:supplementary-structural-diversity}, and \ref{fig:supplementary-ensemble-diversity}).

For these two additional molecules, conformers were generated starting from SMILES after adding hydrogens. We generated 50 conformers per molecule with RDKit 2024.03.3 using ETKDGv3~\cite{Wang2020_ETKDG} (\texttt{pruneRmsThresh=0.01}, \texttt{maxAttempts=5}, \texttt{useRandomCoords=False}), pre-optimized them with MMFF94 where the force field was applicable, and removed duplicates at a heavy-atom best-RMSD threshold of 0.1~\AA, retaining the 10 lowest-energy geometries. Each retained geometry was optimized with xTB 6.7.1 (GFN2, \texttt{--opt tight}, charge 0), and the lowest-energy xTB structure was passed to CREST 3.0.2 conformer search~\cite{Pracht2024_CREST} (GFN2, charge 0) to produce the final ensemble. The 30 lowest-energy CREST conformers were exported as individual XYZ files with their xTB relative energies. Each exported conformer was then evaluated by single-point r\textsuperscript{2}SCAN-3c calculations with ORCA 6.1.1~\cite{Neese2025orca6}; these r\textsuperscript{2}SCAN-3c relative energies served as the scoring reference. At this level, 250 conformer pairs for the macrocycle and 104 for the hydroperoxide satisfy $|\Delta E|\ge2.0$~kcal~mol$^{-1}$.

\subsection*{Reference energies and force-field baselines}

The r\textsuperscript{2}SCAN-3c single-point calculations~\cite{Grimme2021_r2SCAN3c} were performed with ORCA version 6.1.1~\cite{Neese2025orca6} on the same fixed geometries shown to the models. The molecular identities and internal geometries were identical to those in the source GEOM ensembles; only the numerical Cartesian coordinates in the XYZ text differed because each conformer had been rigidly rotated. No geometry reoptimization, solvent model, zero-point correction, or thermal free-energy correction was applied. Within each molecule, energies were shifted so that the lowest conformer had $\Delta E=0$.

To contextualize model performance, UFF~\cite{Rappe1992}, GFN-FF~\cite{Spicher2020_gfnff}, and MMFF94~\cite{Halgren1996_MMFF94} were evaluated on the same exported ensembles and used as physics baselines. Because the RDKit implementations of UFF and MMFF94 used here require a bonded topology, RDKit 2024.03.3~\cite{rdkit} was applied to extract the connectivity and bond orders encoded in the GEOM-QM9 SMILES string~\cite{Weininger1988smiles} associated with each ensemble and assign them to every conformer of that molecule. We deliberately retained this source connectivity instead of asking RDKit to infer bonds from the Cartesian coordinates, thereby preventing possible topology-assignment errors from affecting the UFF and MMFF94 rankings. GFN-FF single-point energies were calculated with xTB 6.7.1~\cite{xtb}, which determines topology directly from the Cartesian coordinates and does not use SMILES bond orders. Thus, three baselines used the same geometries, while UFF and MMFF94 additionally received the molecular connectivity required by these force fields. A random ranking has expected $CRA=0$. The original GEOM GFN2-xTB relative energies~\cite{Bannwarth2019} were retained only for the reference-sensitivity analysis and were not used as the primary scoring target.

\subsection*{Prompt construction and model evaluation}

Each prompt contained the 30 conformers of one molecule in standard element-labeled XYZ text format. Cartesian coordinates were supplied to eight decimal places. Across the 30 molecular ensembles, prompts averaged approximately 29,000 characters and ranged from approximately 24,000 to 34,000 characters (approximately 58,000 for the macrocycle); the corresponding token counts varied across models because their tokenizers differed. Atom order was strictly identical across all 30 conformers within an ensemble. The user message asked for a complete ordering from most stable to least stable using geometric features such as ring strain, eclipsing interactions, nonbonded contacts, hydrogen-bond geometry, and dipole alignment. We included these examples to define an approximate domain of chemically relevant reasoning and indicate the types of geometric effects the model could consider; they did not reveal which effects were present in a particular molecule or how its conformers should be ordered. Molecule names, SMILES strings, energies, and other ranking-relevant information were not supplied. The conformers were numbered from 1 to 30 solely to allow the model to report a ranking; the prompt explicitly stated that these labels conveyed no information about the correct energetic order. The exact text of the prompt was as follows (the label lists “1, 2, ..., 30” are abridged):

\begin{samepage}
\noindent\textbf{System prompt:}\par
\begin{lstlisting}[style=prompt]
You are an expert in computational chemistry.
\end{lstlisting}

\noindent\textbf{User prompt for conformer-ranking evaluation:}\par
\begin{lstlisting}[style=prompt]
You are given 30 conformers of the same molecule in XYZ format (same connectivity, different coordinates).

Each structure has a fixed numeric label: 1, 2, ..., 30 in the headers [Conformer 1] ... [Conformer 30]. Labels are arbitrary identifiers -- do not assume that smaller or larger numbers, or numeric order, says anything about stability. The order of blocks below is random and MUST NOT be used as a hint for relative stability.

There are no thermodynamic quantities in the files. Your task is to rank the 30 structures by estimated relative stability (most stable first, least stable last) using reasonable chemical and steric arguments based on the coordinates (e.g., ring strain, eclipsing interactions, close nonbonded contacts, hydrogen-bond donor/acceptor proximity, dipole alignment). If differences are subtle, still output a full ranking and say what is uncertain.

Task: Rank from most stable to least stable. Keep the explanation tied to geometry.

Output: a single JSON object only-no markdown fences, no text before or after.
Because generation is sequential, put "reasoning" FIRST (step-by-step analysis of geometry, strain, contacts, uncertainty), then "ranking" LAST so the permutation follows your analysis rather than rationalizing a ranking already written.

Schema (use exactly 30 entries in "ranking", each label from 1 to 30 exactly once). Use JSON **strings** for labels, e.g. "3" not bare numbers:
{
  "reasoning": "Analyze hydrogen bonds, steric clashes, ring strain, etc. step-by-step BEFORE committing to the final order.",
  "ranking": ["1", "30", "..."]
}

Rules: "ranking" must be a permutation of {1, 2, ..., 30} (only these numeric string tokens, not the words "Conformer ...").

[Conformer k]
<XYZ block>
\end{lstlisting}
\end{samepage}

Because LLM performance can depend on where relevant information appears within a long context~\cite{liu-etal-2024-lost}, three independently shuffled presentations were generated for each molecule using seeds 42, 43, and 44. The conformer labels and the order in which their coordinates appeared were randomized, separating geometry ranking from positional bias, and the same three presentations were reused for every model. One response was collected for each model, molecule, and presentation.

Finally, we used OpenRouter as a model aggregator, standardizing inference settings wherever provider interfaces allowed while leaving each provider’s default parameters unchanged. Sampling temperature was fixed at 1.0 and was not tuned separately for individual models. Constructing a complete ranking of 30 structures is a multistep task, so this setting allowed some variation in the model’s reasoning rather than forcing deterministic decoding. Research on reasoning models shows that low temperature and greedy decoding can induce looping and repetitive reasoning chains, whereas higher temperature reduces looping by promoting exploration~\cite{pipis2025loop}. We therefore used temperature 1.0 as a common untuned setting and scored each ranking independently; the three run-specific scores were then averaged for each molecule. For GPT-6 Astra, GPT-5.6 Sol, GPT-5, and o4-mini, the provider schemas did not expose a temperature parameter, so the requests used provider-defined decoding behavior. Reasoning configurations and parameter counts are listed in Supplementary Table~\ref{tab:supplementary-reasoning}; the remaining shared inference settings are in Supplementary Table~\ref{tab:supplementary-inference-settings}.

\subsection*{Output validation and resumed runs}

Models were prompted to return a complete permutation of the 30 conformer labels, and outputs were parsed programmatically. API connection timeouts and truncated text responses were re-queried, and completed cases were retained when a run was resumed. Scoring used complete, valid permutations of all 30 labels.

\subsection*{Energy-recall test}

We also conducted an energy-recall test of whether models could recover molecule-specific energetic information from training. For each of the 25 GEOM-QM9 benchmark molecules, we randomly selected one of its 30 conformers using a fixed seed of 42 and used the selected geometry across all evaluated models. Energy annotations were removed from the coordinate input, leaving only atom identities and Cartesian coordinates; no molecular name or SMILES string was supplied. This test used a single response per model and molecule.

For this task, we asked 13 models (11 open-weight models, GPT-6 Astra, and Claude Opus 5) to return a single non-negative estimate of the conformer’s energy above the global ensemble minimum of the same molecule, in kcal~mol$^{-1}$. The prompt explicitly encouraged molecular recognition and use of information recalled from training, with steric and conformational reasoning as a fallback. We compared each estimate with the selected conformer’s reference relative energy (from original GEOM-QM9) and calculated the mean absolute error across the 25 molecules. The exact prompt is presented below, and results are provided in Extended Data Fig.~\ref{fig:supplementary-energy-recall}:

\begin{samepage}
\begin{lstlisting}[style=prompt]
Below are Cartesian coordinates of ONE gas-phase conformer (no SMILES, name, or energy).

Estimate how many kcal/mol HIGHER this conformer is than the global minimum of the same molecule.

Use whatever you can:
1. Infer the molecular formula from the atom list.
2. Reconstruct connectivity / bonding from the geometry.
3. If you recognize the compound or recall similar molecules from your training data (GEOM, QM9, literature, etc.), use that knowledge.
4. If you cannot identify it, estimate from sterics and conformational reasoning.

Reply with one non-negative number in kcal/mol. Nothing else.
\end{lstlisting}
\end{samepage}

\subsection*{Explanation-text concepts quantification, LOESS trend fitting and geometric H-bond definition}

For every explanation (one model $\times$ molecule $\times$ seed) we counted occurrences of a fixed, case-insensitive chemical-concept dictionary—H-bonding; steric clash/crowding; ring strain; torsion/eclipsing; dipole/electrostatics; dispersion/van der Waals; hyperconjugation/anomeric effects; conjugation/aromaticity; nonbonded/transannular interactions; 1,3-diaxial interactions; flexibility/entropy; quantitative geometry; solvation/environment; and symmetry—and recorded the total number of words in the same text. H-bond vocabulary included hydrogen bond/hydrogen-bond, H-bond/H bond/Hbond, donor, acceptor, and $\mathrm{H}\cdots\mathrm{O}$, $\mathrm{H}\cdots\mathrm{N}$ and reverse H-bond contacts. To relate language usage to accuracy, counts were converted to relative abundances $X = n_{\mathrm{concept}}/\mathrm{words}$, $\pm3\sigma$ outliers were removed within each model, and $X$ was rescaled to the model’s own mean. We then fitted, independently for each model, a smooth locally weighted curve (LOESS) of conformer-ranking accuracy $\tau$ against $X$. The curve used tricube-weighted local quadratics over a wide bandwidth (so it is nearly a globally weighted fit), was evaluated across the model’s $X$ range with seeds pooled, and is shown with a 95\% confidence band.

Conformer pairs were classified by H-bond involvement for Fig.~\ref{fig:ranking-heatmap}a--c and the fair-pair analyses in Fig.~\ref{fig:ranking-heatmap}d--f. A conformer was labeled hydrogen-bonded when any hydrogen bound to O or N lay within 2.5~\AA of a second O or N with a donor–H···acceptor angle of at least $120^\circ$, evaluated on the fixed geometries used for the r\textsuperscript{2}SCAN-3c reference energies. Among the 3,802 pairs with $|\Delta E| \ge 2.0$~kcal~mol$^{-1}$, 2,739 had no H-bond on either side, 258 had equal H-bond counts on both sides, and 805 had different counts: 715 were conventional pairs in which the H-bonded conformer is lower in energy, and 90 were unconventional pairs in which it is higher. The 2,739 pairs with no H-bonds and the 258 equal-count pairs together make up the 2,997 equal-hydrogen-bond pairs used in Fig.~\ref{fig:ranking-heatmap}d--f. As a non-energetic reference we scored a hydrogen-bond counting rule (HBC): within each pair, the conformer with more intramolecular hydrogen bonds was predicted to be more stable, and pairs with equal counts were counted as half-correct. Its CRA was computed on the same pairs as for the force fields and averaged over molecules; by construction, the rule scores $+1$ on conventional pairs, $-1$ on unconventional pairs, and $0$ on pairs with equal hydrogen-bond counts (no-H-bond and fair pairs).

\subsection*{MMFF94 energy-term decomposition for Fig.~\ref{fig:benchmark}c}

For every conformer pair with $|\Delta E_{\mathrm{r}^2\textrm{SCAN-3c}}|\ge2.0$~kcal~mol$^{-1}$,
we decomposed the MMFF94 energy difference into its seven components (bond, angle,
stretch--bend, torsion, out-of-plane, van der Waals, and electrostatic) with Open Babel,
using the same SMILES-derived topology and the fixed geometries as the MMFF94 baseline.
For a pair of conformers labelled 1 and 2, each term was oriented along the
r\textsuperscript{2}SCAN-3c ordering as $\Delta E_t=[E_t(1)-E_t(2)]\,\mathrm{sign}(\Delta E_{\mathrm{r}^2\textrm{SCAN-3c}})$, where $\Delta E_{\mathrm{r}^2\textrm{SCAN-3c}}=E_{\mathrm{r}^2\textrm{SCAN-3c}}(1)-E_{\mathrm{r}^2\textrm{SCAN-3c}}(2)$. Thus, positive $\Delta E_t$ indicates that the term favours the conformer with lower r\textsuperscript{2}SCAN-3c energy. Pairwise agreement between the two methods is shown molecule by molecule in Extended Data Fig.~\ref{fig:supplementary-mmff-alignment}.

Term weights were fitted only on the eight molecules whose total MMFF94 energy
differences reproduced the r\textsuperscript{2}SCAN-3c pairwise energy differences with $R^2\ge0.8$
(molecules 1, 4, 6, 11, 19, 22, 24 and 26; 1,391 pairs). All pairs entered a single fit
per model, without separating them by H-bond class. For each LLM separately, we fitted
pairwise ranking correctness as $P(\mathrm{correct})=\sigma\!\left(b_m+\sum_t c_t\Delta E_t\right)$, with one intercept $b_m$ per molecule and non-negative term coefficients $c_t\ge0$. The three independently shuffled runs entered as separate observations, giving 4,173 observations per model. The fit used L-BFGS-B with an $L_2$ penalty of $10^{-2}$ on the term coefficients. For display,
the coefficients of each model were normalised to sum to one, so each bar gives the share
of the ranking decision attributed to that term; the numbers below the axis give the mean
absolute change of each term, $\overline{|\Delta E_t|}$, over the same 1,391 pairs (each of the eight molecules is weighted equally).
The panel shows the 14 of 28 models whose correct-ranking share, averaged over molecules on
the reliable pairs, exceeds that of UFF evaluated on the same pairs (UFF: CRA=0.23).

\subsection*{Ranking metrics and aggregation}

For a predicted ranking, let each unordered conformer pair be concordant if the predicted order matches the r\textsuperscript{2}SCAN-3c $\Delta E$ order, and discordant otherwise. Let the reliable-pair Kendall $\tau$ at the threshold $\delta$ be
\[
\tau_{\delta}=\frac{C_{\delta}-D_{\delta}}{C_{\delta}+D_{\delta}},
\]
where $C_{\delta}$ and $D_{\delta}$ denote the counts of concordant and discordant pairs, respectively, among those with $|\Delta E|\ge\delta$. Let the primary metric be computed using conformer pairs with $|\Delta E|\ge2.0$~kcal~mol$^{-1}$.

Because the three presentations probed the same underlying molecular ensemble, each response first yielded one metric value and the three run-specific values were then averaged, producing up to 27 molecule-level scores per model (26 for Qwen 3 32B and 25 for Qwen 2.5 32B). All subsequent summaries used these molecule-level scores as the unit of analysis.

These molecule-level scores then formed the basis of the reported summaries. Let $x_1,\ldots,x_n$ be the scores of one model, where $n$ is the number of molecules scored for that model. The reported mean, median, and sample standard deviation in Supplementary Tables~\ref{tab:model-performance-all} and~\ref{tab:model-performance} are
\[
\bar x=\frac{1}{n}\sum_{i=1}^{n}x_i,\qquad
s=\sqrt{\frac{1}{n-1}\sum_{i=1}^{n}(x_i-\bar x)^2}.
\]
Error bars in Fig.~\ref{fig:benchmark}b are $95\%$ Student-$t$ confidence intervals for the mean across molecules,
\[
\bar x \pm t_{0.975,\,n-1}\,\frac{s}{\sqrt{n}},
\]
where $t_{0.975,\,n-1}$ is the $0.975$ quantile of the $t$ distribution
with $n-1$ degrees of freedom ($26$ for the full 27-molecule set,
$25$ for Qwen 3 32B with $n=26$, and $24$ for Qwen 2.5 32B with $n=25$;
$t_{0.975,26}\approx2.056$, $t_{0.975,25}\approx2.060$,
$t_{0.975,24}\approx2.064$).

Extended Data Fig.~\ref{fig:supplementary-molecule-performance}b displays the molecule-level values used in these summaries. For each LLM-molecule pair, the heatmap cell is the mean $\tau$ across the three runs; UFF, MMFF94, and GFN-FF contribute one deterministic value per molecule. Summary statistics for the complete evaluated model cohort are reported in Supplementary Table~\ref{tab:model-performance}.

\paragraph*{Paired comparisons with UFF}

For paired comparisons with UFF, we defined $d_i = \tau_i^{\mathrm{LLM}}-\tau_i^{\mathrm{UFF}}$ for each molecule, using the same molecule-level $\tau$ values as above (i.e., molecules with complete answers across the three runs). We report the mean paired difference with a Student-$t$ 95\% confidence interval constructed exactly as for the Fig.~\ref{fig:benchmark}b mean intervals (full cohort: $n=27$, $df=26$; Qwen 3 32B: $n=26$, $df=25$; Qwen 2.5 32B: $n=25$, $df=24$, because they did not provide complete answers for one and two additional molecules, respectively). Significance against no mean difference was assessed with a two-sided one-sample $t$-test of $\{d_i\}$ against zero, equivalently a paired $t$-test of LLM and UFF molecule-level scores, with $df=n-1$ (26 for the full 27-molecule set, 25 for Qwen 3 32B, and 24 for Qwen 2.5 32B). As supporting sensitivity analyses, we also report a two-sided paired (sign-flip) permutation test and a two-sided Wilcoxon signed-rank test of the molecule-level differences. A model was described as showing a significant paired advantage over UFF only when the Student-$t$ 95\% interval excluded zero; the permutation and Wilcoxon tests were not used to define this criterion. Full results for every model are given in Supplementary Table~\ref{tab:supplementary-uff-comparisons}.

\paragraph*{Association with release date}

Across the 28 models, mean CRA across the 2,997 equal-HB (fair) pairs was related to calendar release date. Each release date was taken from the curated registry reported in Supplementary Table~\ref{tab:supplementary-model-identities} and converted to a numerical abscissa of fractional years since 1 January 2024, i.e., \texttt{(date - 2024-01-01).days / 365.25}. Ordinary least-squares (OLS) regression of mean $\tau$ on the same abscissa yielded the coefficient of determination $R^2$ and the two-sided $p$-value for the fit ($df=n-2=26$). Both are reported in Extended Data Fig.~\ref{fig:supplementary-model-correlates}.

\paragraph*{Best-to-date and worst-to-date lines}

For the release-date panels, models were sorted by release date and the plotted per-model mean CRA values were accumulated as a running maximum (black lines in Fig.~\ref{fig:benchmark}b and Fig.~\ref{fig:ranking-heatmap}a,b) or a running minimum (red line in Fig.~\ref{fig:ranking-heatmap}c). The resulting step function changes only when a newly released model exceeds all earlier values (running maximum) or falls below all earlier values (running minimum). Models released within a short window are offset horizontally for legibility, and the step line follows the same offsets. The line is a descriptive summary of the cohort, not a fitted trend, and no confidence interval is attached to it.

\paragraph*{Cross-model regressions}

External benchmark scores (GPQA Diamond, SciCode, ARC-AGI-1, and ARC-AGI-2) were taken from a frozen snapshot of the OpenRouter rankings and the Modelgrep API; for each model and benchmark the higher of the two published values was retained. The exact values used are included in the accompanying data. These external scores may be updated by their providers over time, so live values may differ from those used here.

Extended Data Figs.~\ref{fig:supplementary-model-correlates} and~\ref{fig:supplementary-accuracy-cost-benchmarks} relate each model’s mean CRA across 2,997 equal-HB (fair) pairs to an external predictor by OLS, with one point per model. In each panel, $R^2$ is the squared Pearson correlation of the fitted line, and the reported $p$-value is the two-sided test that the OLS slope equals zero. Display jitter used in the figures did not enter the fits.

\emph{Parameter count ($n=16$).} Models with unpublished total parameter counts were omitted. The predictor was $\log_{10}$ of total parameters in billions; the OLS fit used this log-transformed abscissa rather than raw parameter count.

\emph{Total API cost ($n=17$).} For each model, total API cost was calculated by summing the per-call \texttt{cost\_total} fields over only those molecule-seed combinations for which a valid ranking (and hence a fair-pair CRA) was obtained. Only models for which this total could be computed from the recorded cost fields were included. The OLS fit used the raw dollar total without log transformation.

Total API cost was associated with conformer-ranking accuracy on the 2,997 fair pairs across the 17 matched models ($R^2=0.82$, $p<0.001$; Extended Data Fig.~\ref{fig:supplementary-accuracy-cost-benchmarks}a). At the low-cost end, DeepSeek V4 Flash 0731 achieved CRA=0.23 for \$1.20 over the full benchmark above GPT-5 (CRA=0.19, \$18.41), but below UFF (CRA=0.42). At intermediate cost, Gemini 3.6 Flash reached CRA=0.34 for \$26.35, compared with GPT-5.6 Sol at CRA=0.30 for \$48.69. GPT-6 Astra achieved CRA=0.66 for \$103.23: 3.92 times the expenditure of Gemini 3.6 Flash for an increase of 0.32 in CRA. These are descriptive comparisons at the recorded provider prices, not estimates of the effect of increasing a fixed model’s reasoning budget. API expenditure combines input and output usage with provider-specific pricing and cannot be equated to a common amount of computation.

\emph{GPQA Diamond ($n=25$).} External scores were taken from OpenRouter rankings and the Artificial Analysis fields served by Modelgrep for the matched model identifiers, retaining the higher published value for each benchmark field. Models without a published score in either source were omitted. Predictors were used on the reported $[0,1]$ scale without further transformation.

\emph{SciCode ($n=20$).} External scores were obtained using the same OpenRouter-Modelgrep matching and higher-value rule. Models without a published score in either source were omitted, and predictors were used on the reported $[0,1]$ scale without further transformation. The fit is shown in Extended Data Fig.~\ref{fig:supplementary-accuracy-cost-benchmarks}c.

\emph{Qwen-declustered comparison.} For GPQA, a second OLS fit excluded the four Qwen variants forming the distinct cluster—Qwen 3.5 27B, Qwen 3.6 27B, Qwen 3.5 35B A3B, and Qwen 3.6 35B A3B—leaving $n=21$.

Four Qwen 3.5 and 3.6 variants lie below the overall association between GPQA Diamond and conformer-ranking accuracy. Excluding this cluster changes the fit from $R^2=0.55$ ($n=25$) to $R^2=0.61$ ($n=21$; both $p<0.001$; Extended Data Fig.~\ref{fig:supplementary-accuracy-cost-benchmarks}b). Because this exclusion is motivated by the observed cluster, it is a descriptive sensitivity analysis rather than independent confirmation of a stronger relationship. Qwen 3.8 27B lies closer to the overall trend, showing that the deviation is not shared by every evaluated member of the family.

The cluster includes sparse 35B A3B models, with approximately 3B active parameters per token, and dense 27.8B models. Limited active capacity and differences in inference behavior are possible explanations, but the comparisons do not isolate either cause. Across the benchmark tasks, recorded reasoning text for Qwen 3.8 27B contained 8.60 million characters versus 2.09 million for Qwen 3.6 27B, a 4.11-fold difference. Trace length is a measure of recorded output, not a calibrated measure of reasoning compute; the models also differ in their training and development. The association is consistent with a role for inference-time reasoning, but does not establish that additional computation caused the recovery in ranking accuracy.

\emph{ARC-AGI-1 and ARC-AGI-2 ($n=15$ each).} External scores were matched to the evaluated model identifiers using the same OpenRouter-Modelgrep matching and higher-value rule as for GPQA Diamond and SciCode. Models without a published score for the corresponding ARC-AGI benchmark in either source were omitted, leaving the same 15-model cohort for both analyses. Predictors were used on the reported $[0,1]$ scale without further transformation; the ARC-AGI-2 fit is shown in Extended Data Fig.~\ref{fig:supplementary-model-correlates} and the ARC-AGI-1 fit in Extended Data Fig.~\ref{fig:supplementary-accuracy-cost-benchmarks}d.

\subsection*{Coupled-cluster reference validation}

DLPNO-CCSD(T) was used to validate the r\textsuperscript{2}SCAN-3c conformer ordering. The 30 initial molecular ensembles were sorted in ascending order by their r\textsuperscript{2}SCAN-3c energy span, defined as the difference between the highest- and lowest-energy conformers, and ten molecules approximately evenly spaced across this ordering (with a small random jitter) were selected to give a ten-molecule validation subset spanning almost the entire range. One of the ten did not enter the final 25-molecule GEOM-QM9 subset of our benchmark because it contained fewer than 15 conformer pairs satisfying $|\Delta E|\ge2.0$~kcal~mol$^{-1}$; it was nonetheless retained in the DLPNO-CCSD(T) comparison.

For these structures, single-point DLPNO-CCSD(T) calculations~\cite{Riplinger2013} used the unchanged neutral singlet geometries, the def2-TZVP orbital basis~\cite{Weigend2005}, TightPNO thresholds~\cite{Liakos2015}, the RIJCOSX approximation~\cite{Neese2009rijcosx}, and the cc-pVTZ/C auxiliary basis~\cite{Hattig2005} for correlation fitting.

The following ORCA input commands were used to calculate the energies:

\begin{samepage}
\begin{lstlisting}[style=prompt]
! r2SCAN-3c SP
! ENGRAD
%pal nprocs 4 end

* xyzfile 0 1 geometry.xyz
\end{lstlisting}
\end{samepage}

and

\begin{samepage}
\begin{lstlisting}[style=prompt]
! DLPNO-CCSD(T) def2-TZVP RIJCOSX TightPNO cc-pVTZ/C SP
%pal nprocs 16 end
%maxcore 4000

* xyzfile 0 1 geometry.xyz
\end{lstlisting}
\end{samepage}

Using these energies, pairwise agreement between r\textsuperscript{2}SCAN-3c and DLPNO-CCSD(T) was evaluated across all ten validation molecules, including the molecule later excluded from the final benchmark, using the same r\textsuperscript{2}SCAN-3c energy-gap thresholds as the benchmark. There was 100\% agreement on pairs with $|\Delta E|\ge1.5$ and $|\Delta E|\ge2.0$~kcal~mol$^{-1}$ (1,691 and 1,213 pairs satisfying these gaps, respectively). At each threshold $\delta$, a pair was counted as agreeing when the two methods assigned the same relative order.

Finally, model-order sensitivity was assessed by rescoring all 28 LLMs on the nine validation molecules retained in the final 27-molecule set, once against r\textsuperscript{2}SCAN-3c and once against DLPNO-CCSD(T). Agreement of the resulting model rankings was near-perfect at the 1.0 and 1.5~kcal~mol$^{-1}$ thresholds (Spearman’s $\rho=0.9973$ and $0.9945$; Kendall’s $\tau=0.98$ and $0.96$, respectively) and remained high at 2.0~kcal~mol$^{-1}$ ($\rho=0.9650$, $\tau=0.86$).

\subsection*{Model release dates}

Release dates refer to the public model release, not to the date of API access; they were taken from the provider’s official release announcement and cross-checked against the model page on the serving aggregator (OpenRouter) where available. Where sources conflicted, 24 February 2026 was used for both Qwen 3.5 27B (reported as 23 or 24 February 2026) and Qwen 3.5 35B A3B (reported as 24–25 February 2026). The release date for Qwen 3.7 Max was taken from OpenRouter. For Qwen 3.5 Plus, the dated snapshot identifier encodes 20260420, and 20 April 2026 was used as its release date. These same dates are used in Fig.~\ref{fig:benchmark}b and Fig.~\ref{fig:ranking-heatmap}a--c.

\section*{Data Availability}

The data supporting this study, including the benchmark geometries, reference energies, model prompts, model rankings, recorded model explanations and available reasoning traces, and supplementary analysis results, are available at \url{https://github.com/TheorChemGroup/LLMConfBench}.

\section*{Code Availability}

The code supporting this study is available at \url{https://github.com/TheorChemGroup/LLMConfBench} and is released under the MIT License.

\section*{Acknowledgments}

We thank Andrei Vasiliev for project management, organizational coordination, and logistical support throughout this work. Computational resources were provided in part by the HPC facilities at NRU HSE.

\section*{Funding}

Russian Science Foundation grant \#22-73-10124-P (MGM)

\section*{Author Contributions}

Conceptualization: G.A.S., I.V.P., S.N.O., I.V.A., M.G.M.; Methodology: G.A.S., S.N.O.; Investigation: G.A.S., S.N.O.; Data curation: G.A.S.; Formal analysis: G.A.S.; Visualization: G.A.S., T.V.L., S.N.O., M.G.M.; Software: T.V.L.; Writing – original draft: G.A.S., S.N.O.; Writing – review \& editing: I.V.P., S.N.O., I.V.A., M.G.M.; Supervision: M.G.M. All authors reviewed and approved the manuscript.

\section*{Competing Interests}

The authors declare that they have no competing interests.

\clearpage

\begingroup
\captionsetup[figure]{
  font={normalfont,small},
  labelfont=bf,
  justification=justified,
  singlelinecheck=false,
  labelsep=barsep,
  margin=0pt
}
\renewcommand{\figurename}{Extended Data Fig.}
\renewcommand{\thefigure}{\arabic{figure}}
\renewcommand{\theHfigure}{extendeddata.\arabic{figure}}
\setcounter{figure}{0}
\begin{figure}[H]
  \centering
  \includegraphics[width=\textwidth,height=0.82\textheight,keepaspectratio]{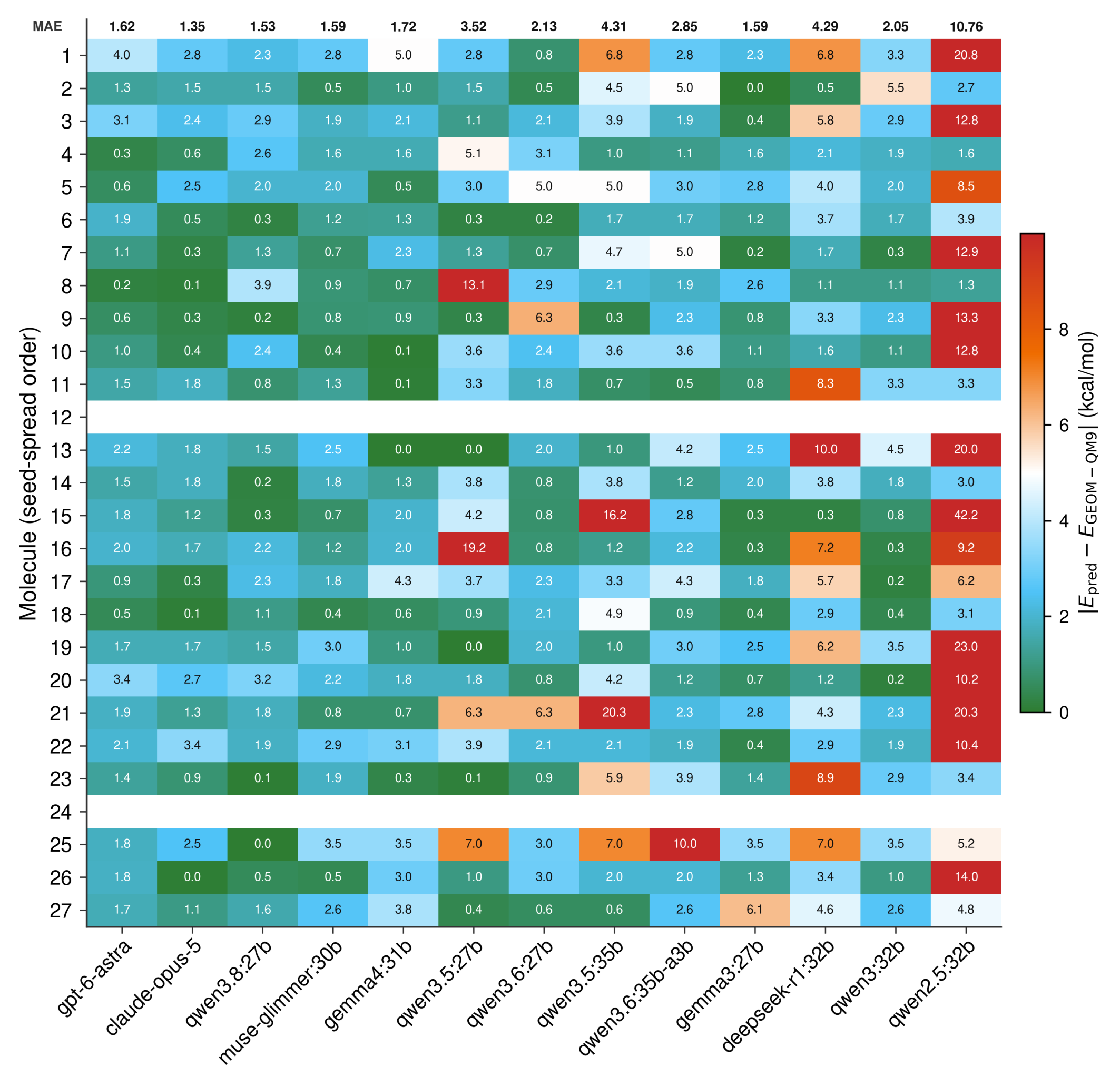}
  \caption{\textbf{Prediction of reference conformer energies as a test of molecular memorization.} Absolute errors between model-predicted relative energies and the original GEOM-QM9 reference values for one selected conformer of each of the 25 benchmark molecules, evaluated across 13 models. Cell values and colors indicate absolute errors in kcal~mol$^{-1}$; mean absolute errors (MAE) across molecules are shown above the columns. Models were explicitly encouraged to recognize the molecules and recall information from training.}
  \label{fig:supplementary-energy-recall}
\end{figure}
\clearpage
\begin{figure}[H]
  \centering
  \includegraphics[width=\textwidth,height=0.82\textheight,keepaspectratio]{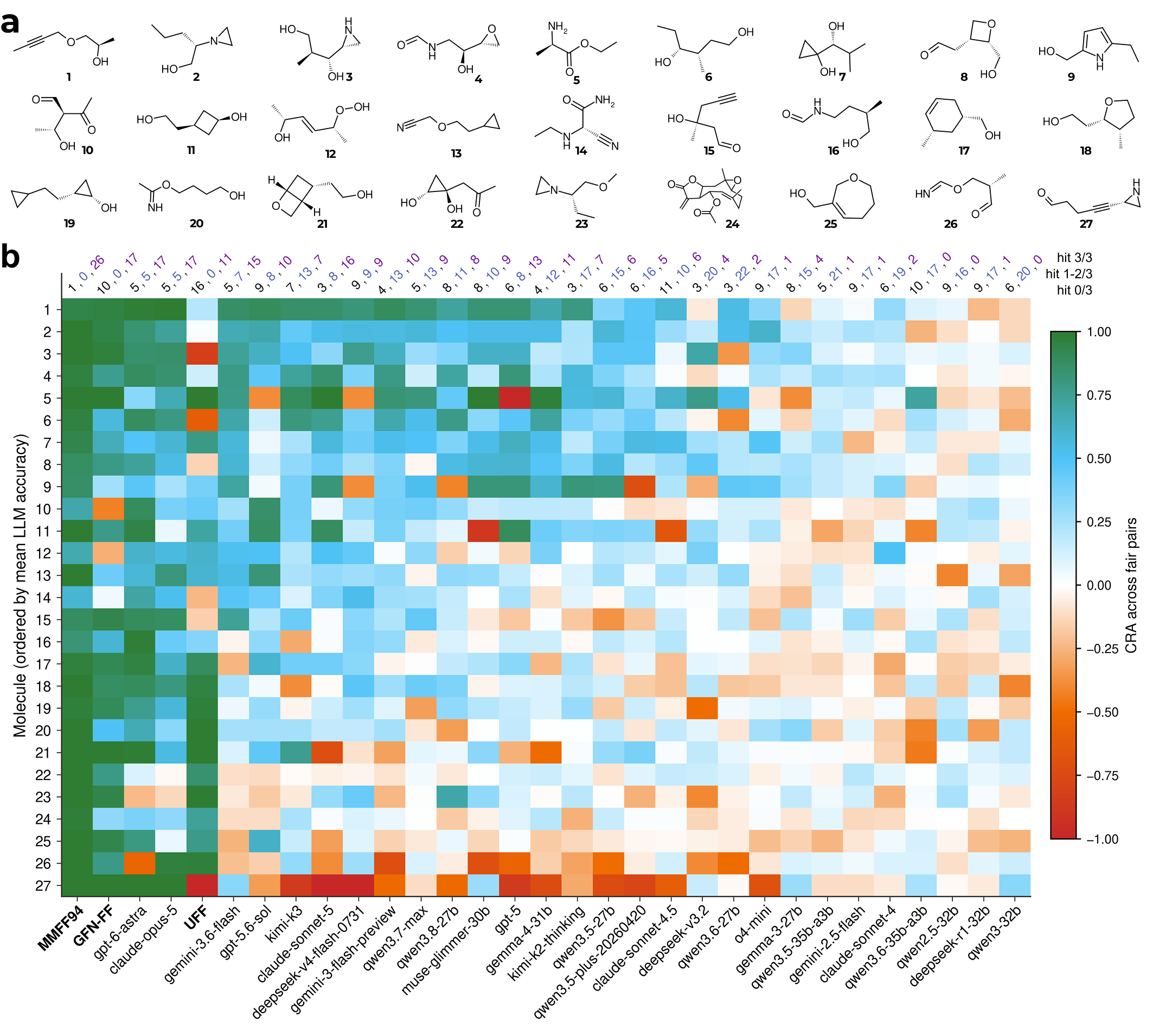}
  \caption{\textbf{Divergence of LLM performance across molecules.} \textbf{A}, Structures of the 25 scored GEOM-QM9 molecules, shown in the same left-to-right order as in all plots, plus two additional out-of-distribution challenge molecules (12 and 24). \textbf{B}, CRA on pairs with equal counts of hydrogen bonds (2,997 pairs). Rows correspond to molecules ordered by mean LLM accuracy; columns correspond to models and force fields, sorted by mean CRA. Cells are colored according to averages across three runs. Counts above columns show recovery of the true energy minimum in the predicted top ten: black, none; indigo, one or two; purple, all three. Force fields are deterministic, so they produce only one ranking per molecule.}
  \label{fig:supplementary-molecule-performance}
\end{figure}
\clearpage
\begin{figure}[H]
  \centering
  \includegraphics[width=\textwidth,height=0.80\textheight,keepaspectratio]{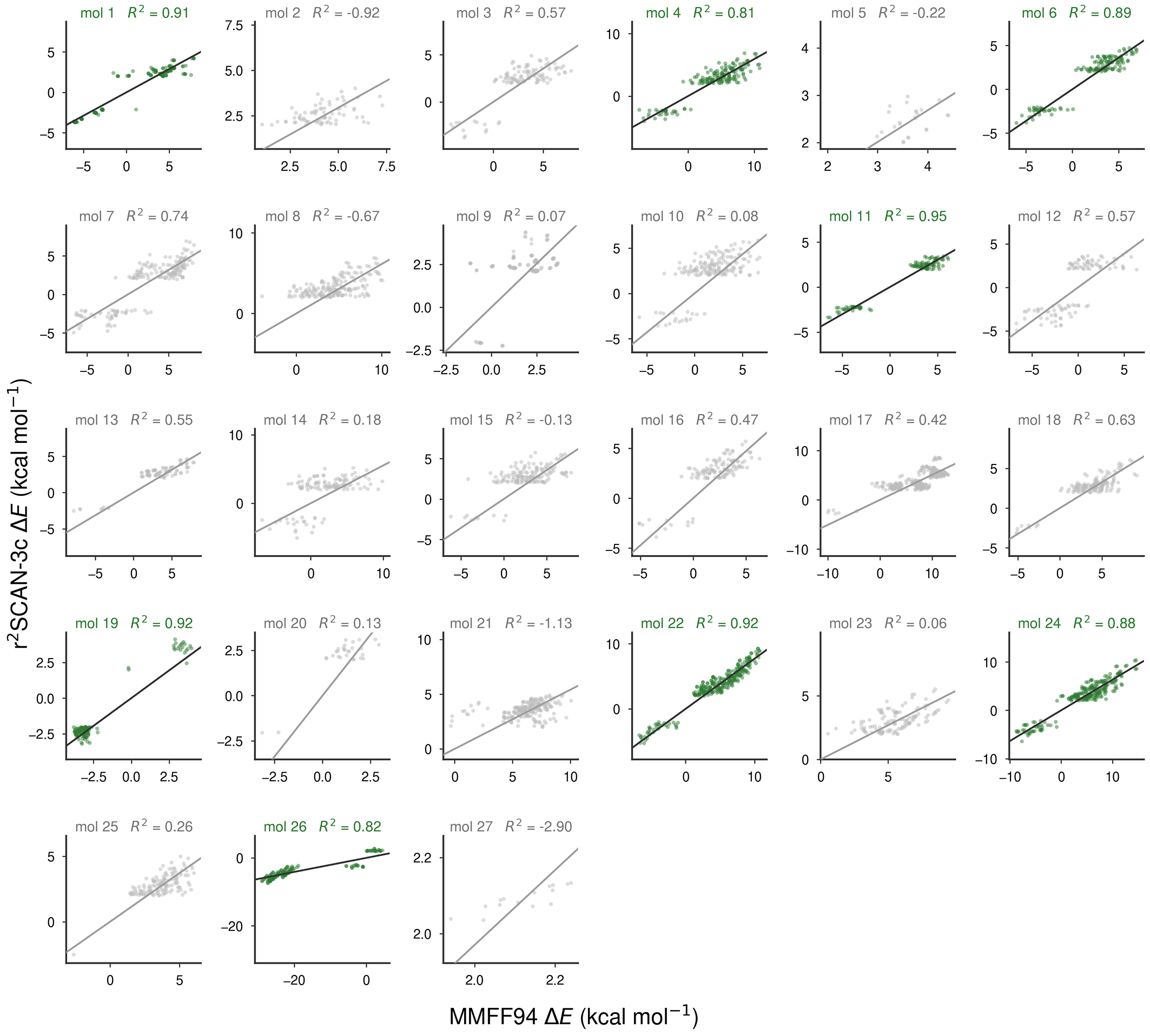}
  \caption{\textbf{Agreement between MMFF94 and r\textsuperscript{2}SCAN-3c conformer-pair energy differences.} Each panel shows pairwise energy differences for one molecule, with $R^2$ computed about the mean for a slope constrained through the origin. Green denotes the eight molecules with $R^2\ge0.8$ used to fit the MMFF94 energy-term model; gray denotes the other 19 molecules, which were excluded from the energy-term fit.}
  \label{fig:supplementary-mmff-alignment}
\end{figure}
\clearpage
\begin{figure}[H]
  \centering
  \includegraphics[width=\textwidth]{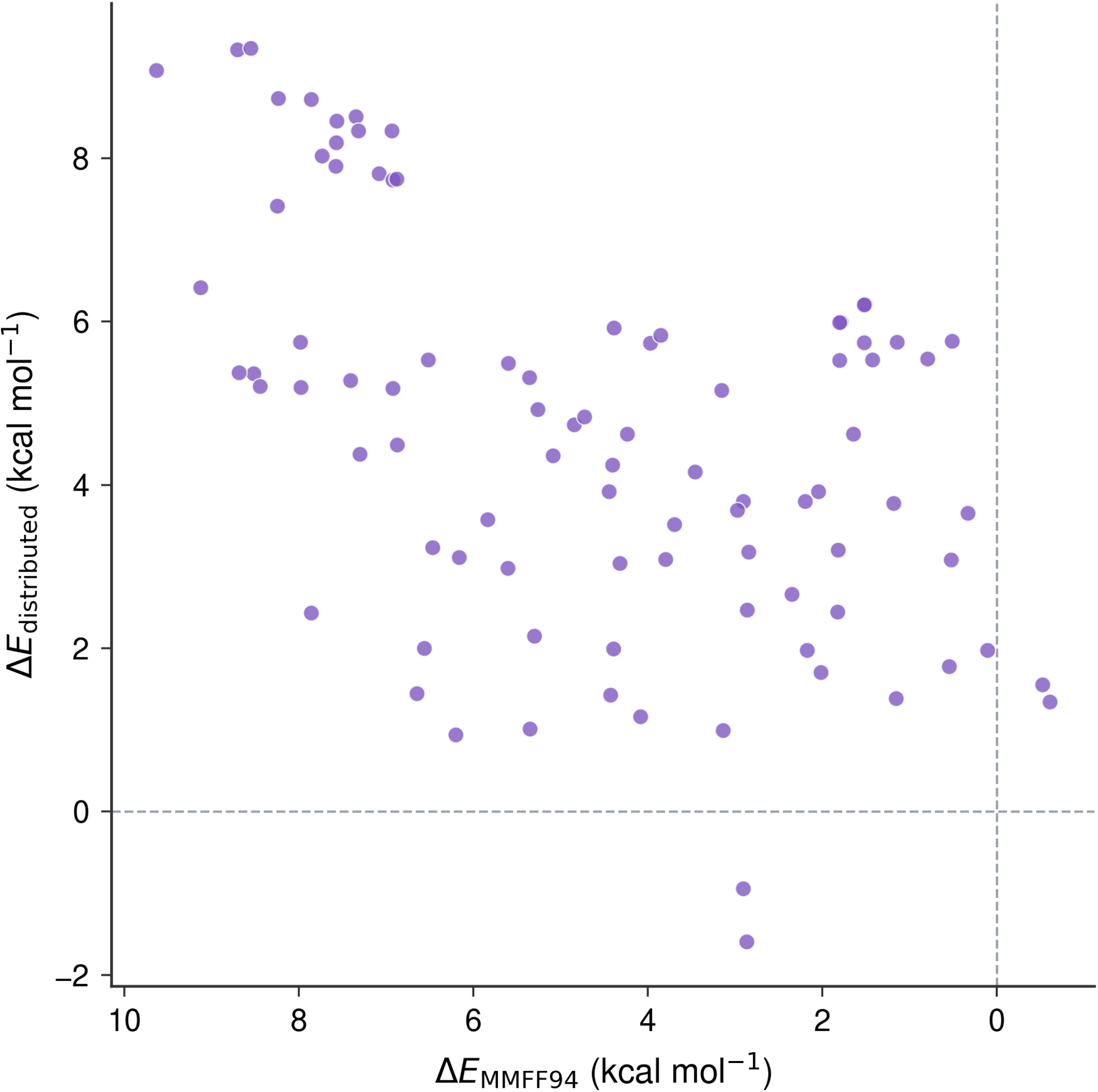}
  \caption{\textbf{Distributed strain versus the MMFF94 energy difference for unconventional pairs.} Each point is one of the 90 pairs in which the more hydrogen-bonded conformer is less stable (11 molecules). Distributed strain is the sum of the Angle, StrBnd, Torsion and OOP MMFF94 terms; both axes are oriented from the lower- to the higher-energy conformer. The x axis is reversed so that zero falls on the right.}
  \label{fig:supplementary-distributed-strain}
\end{figure}

\clearpage
\begin{figure}[H]
  \centering
  \includegraphics[width=\textwidth,height=0.85\textheight,keepaspectratio]{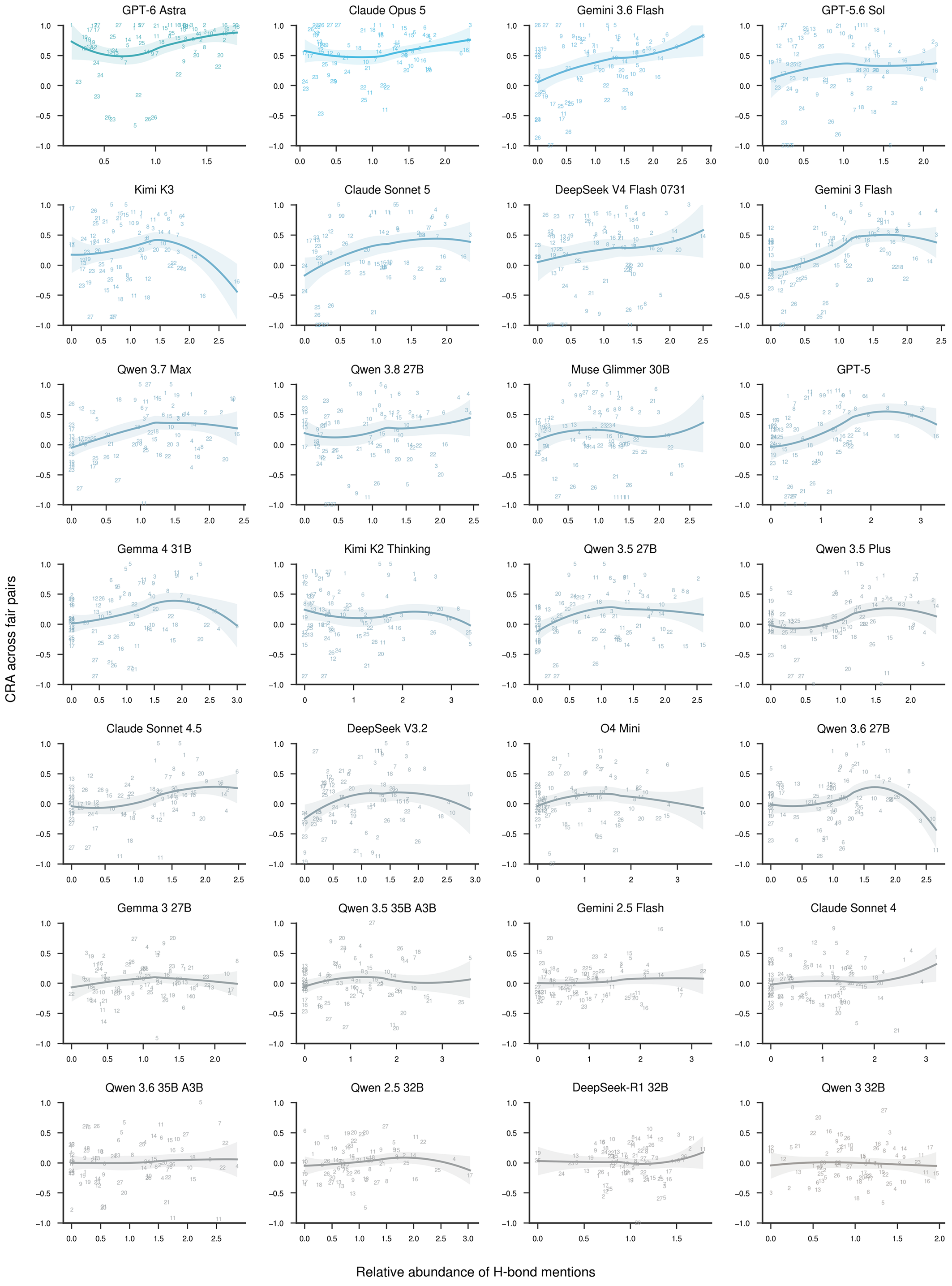}
  \caption{\textbf{Hydrogen-bond mentions as a marker of reliable conformer ranking for all models.} Lines are rolling averages (LOESS fits) with 95\% confidence intervals. For Qwen 3 32B and Qwen 2.5 32B, 77 and 74 explanations, respectively, are plotted out of 78 and 75 available explanations (26 and 25 molecules, respectively, across three seeds).}
  \label{fig:supplementary-hbond-models}
\end{figure}
\clearpage
\begin{figure}[H]
  \centering
  \includegraphics[width=\textwidth]{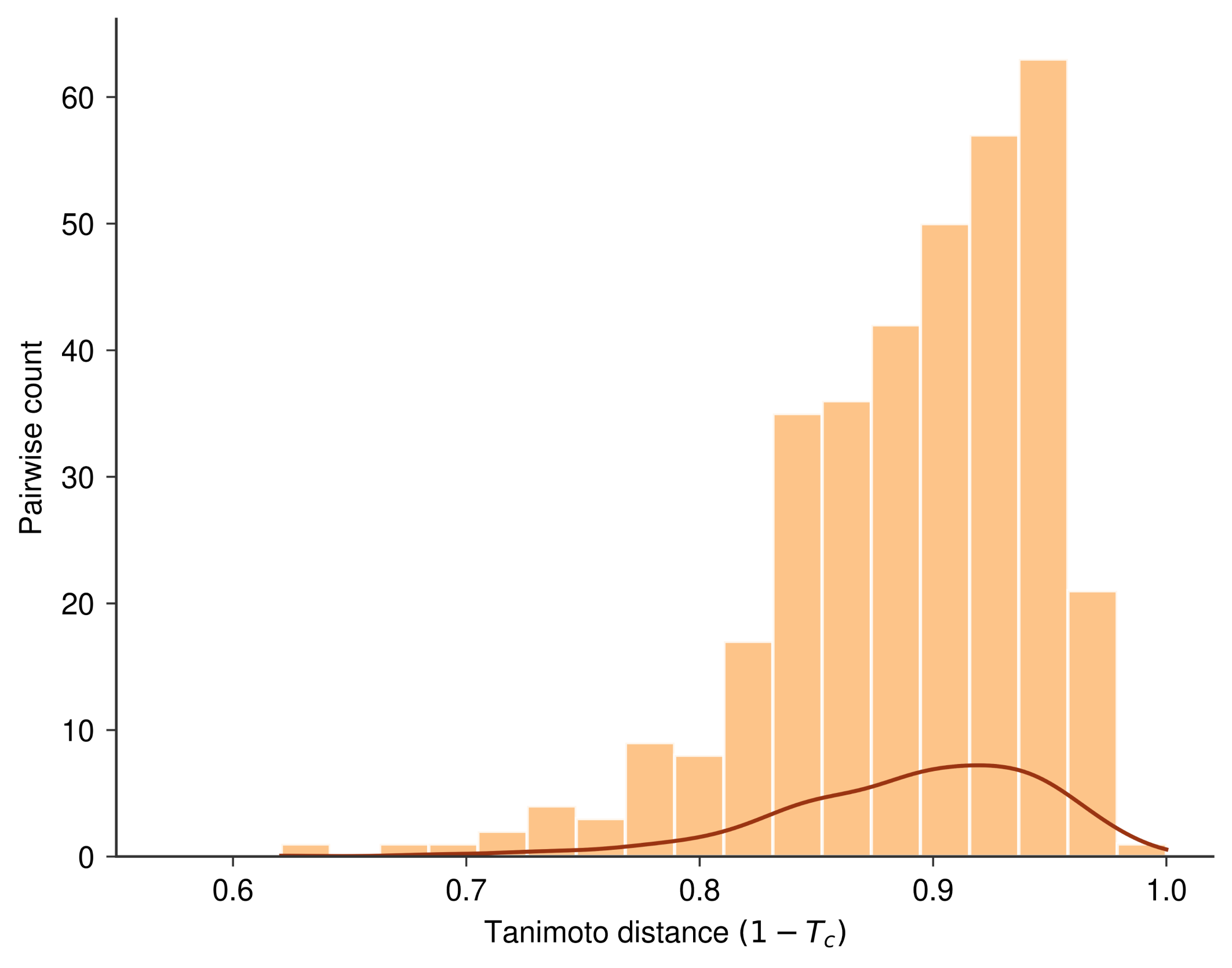}
  \caption{\textbf{Structural diversity of the benchmark and challenge molecules.} Distribution of pairwise Tanimoto distances, $1-T_c$, across all 351 unique molecular pairs. The set comprises the 25 GEOM-QM9 benchmark molecules and two additional challenge molecules absent from QM9: the macrocycle from the study by Kahsai et al. and the peroxide (2R,5R,E)-5-hydroperoxyhex-3-en-2-ol. Larger distances indicate lower molecular similarity. The distribution has a mean of 0.891, a median of 0.897, and a range of 0.621--1.000.}
  \label{fig:supplementary-structural-diversity}
\end{figure}
\clearpage
\begin{figure}[H]
  \centering
  \includegraphics[width=\textwidth,height=0.80\textheight,keepaspectratio]{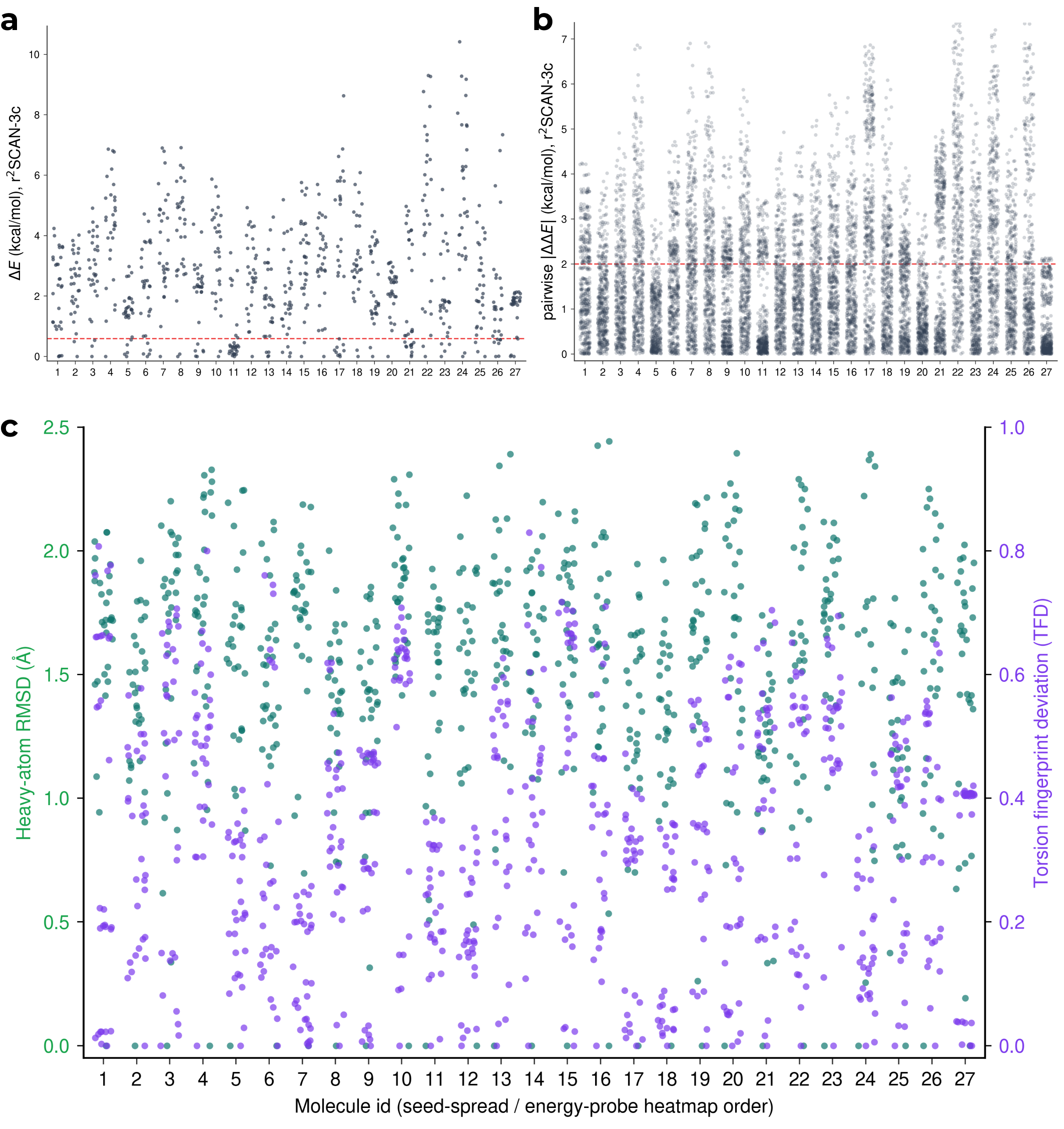}
  \caption{\textbf{Energetic and geometric diversity of the conformer ensembles.} The set comprises the 25 GEOM-QM9 molecules plus two additional challenges---the Kahsai et al. macrocycle and (2R,5R,E)-5-hydroperoxyhex-3-en-2-ol, both absent from QM9. \textbf{A}, r\textsuperscript{2}SCAN-3c conformer energies relative to each ensemble's minimum; the red dashed line marks $k_{\mathrm{B}}T$ at 298~K = 0.592~kcal~mol$^{-1}$. \textbf{B}, Every dot is a pairwise conformer energy difference in the r\textsuperscript{2}SCAN-3c ensembles; the red dashed line marks the 2.0~kcal~mol$^{-1}$ threshold, satisfied by 3,802 of 11,745 pairs. \textbf{C}, Per-conformer heavy-atom RMSD (green, left axis, \AA) and torsion fingerprint deviation (TFD; purple, right axis). Indices \textbf{1--27} are shared between panels and follow the row order in the Extended Data Fig.~\ref{fig:supplementary-molecule-performance}b heatmap.}
  \label{fig:supplementary-ensemble-diversity}
\end{figure}
\clearpage
\begin{figure}[H]
  \centering
  \includegraphics[width=\textwidth,height=0.84\textheight,keepaspectratio]{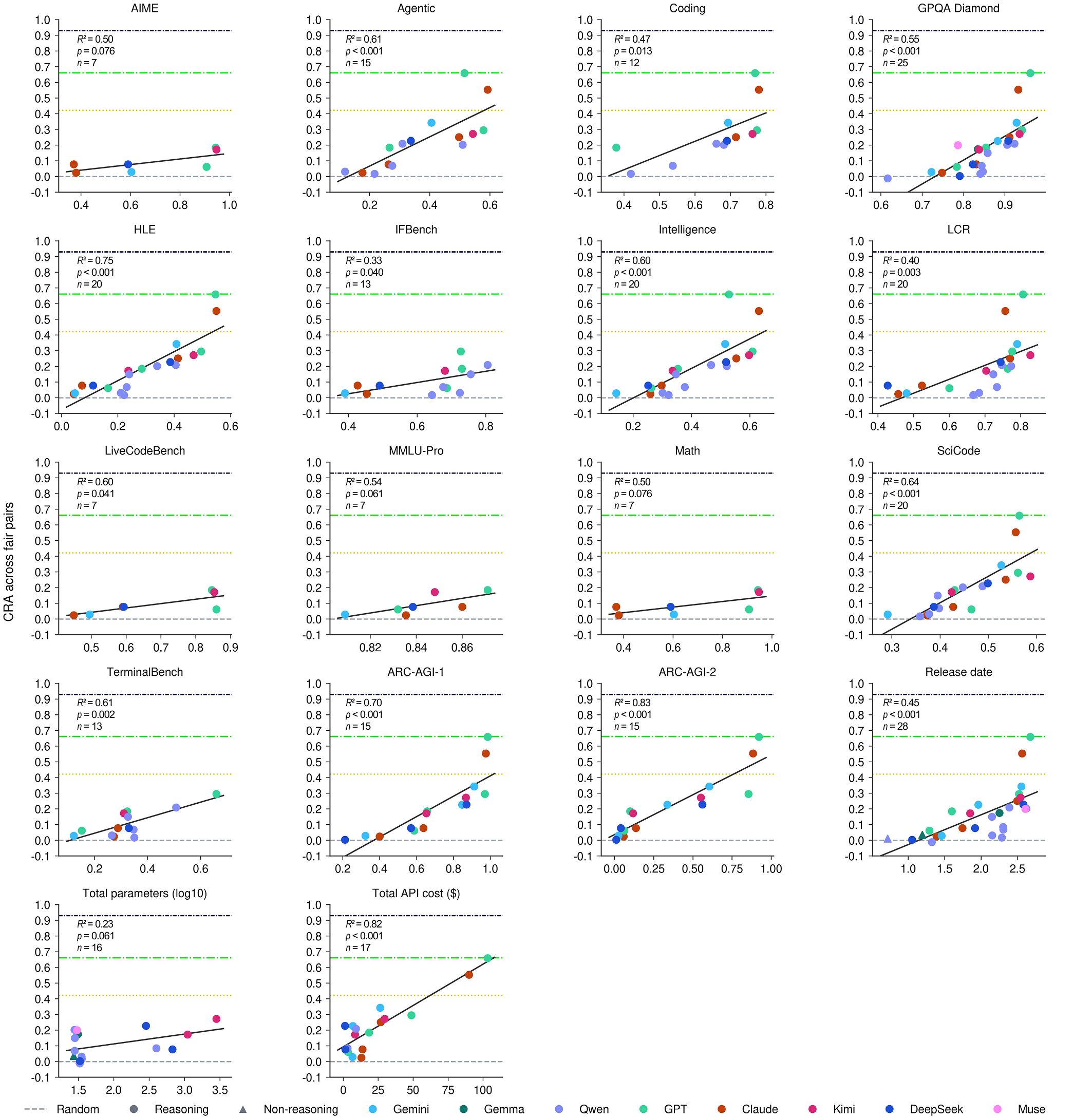}
  \caption{\textbf{Conformer-ranking accuracy versus external benchmarks and model characteristics.} Each panel relates mean CRA across the 2,997 equal-HB (fair) pairs to the predictor named above or below the panel. Solid black lines are least-squares fits; statistics are shown in each panel. Horizontal lines mark random ranking, UFF, GFN-FF, and MMFF94. Colors denote model families, and marker shapes distinguish reasoning and non-reasoning configurations.}
  \label{fig:supplementary-model-correlates}
\end{figure}
\clearpage
\begin{figure}[H]
  \centering
  \includegraphics[width=\textwidth,height=0.84\textheight,keepaspectratio]{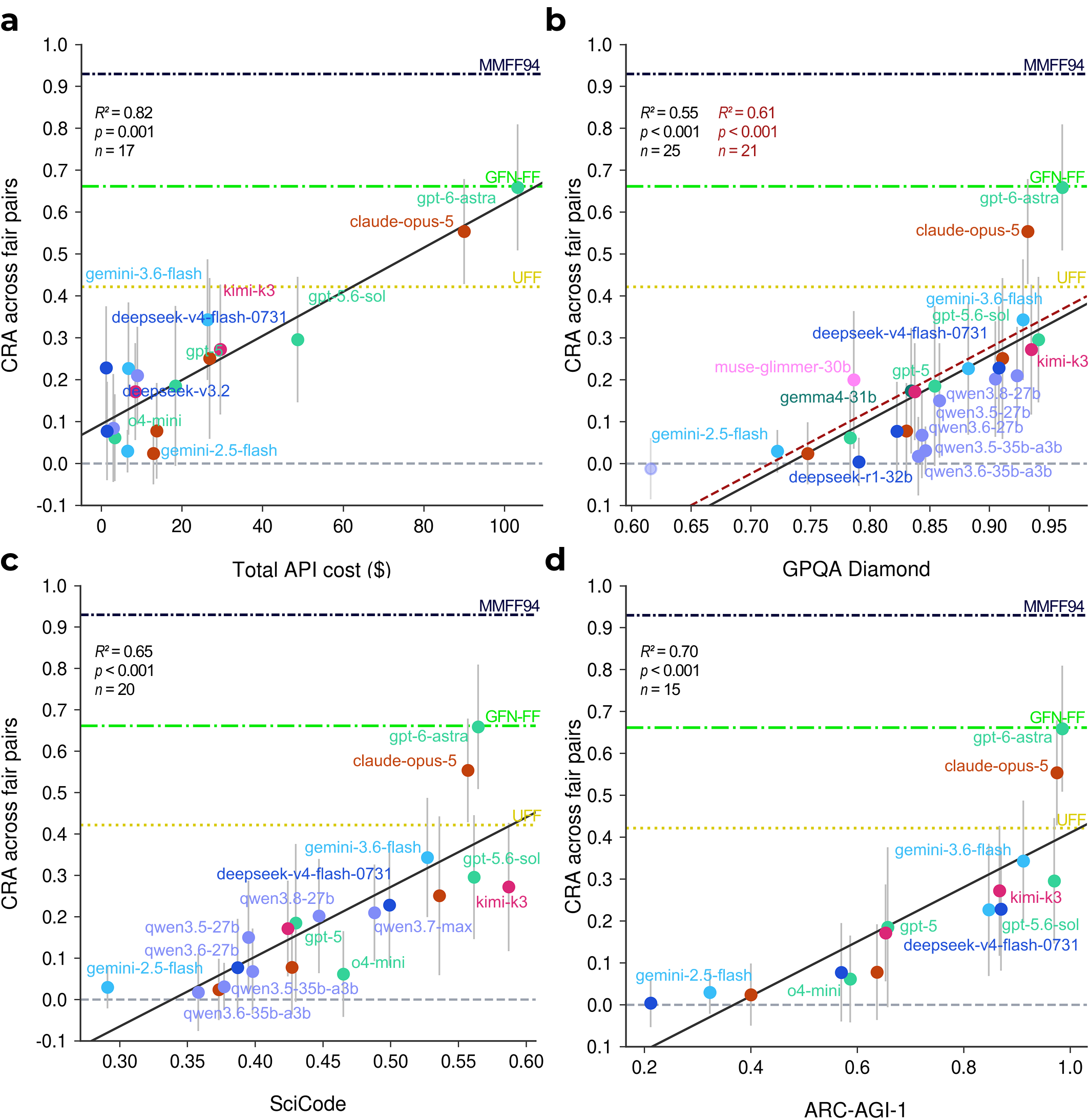}
  \caption{\textbf{Conformer-ranking accuracy versus benchmarks and API cost.} CRA on the 2,997 equal-HB (fair) pairs vs.\ total API cost, GPQA Diamond, SciCode, and ARC-AGI-1. Black lines are least-squares fits (total cost $R^2=0.82$, $n=17$; GPQA $R^2=0.55$, $n=25$, and $R^2=0.61$, $n=21$ excluding the Qwen cluster; SciCode $R^2=0.65$, $n=20$; ARC-AGI-1 $R^2=0.70$, $n=15$); horizontal lines mark random, UFF, GFN-FF, and MMFF94 baselines.}
  \label{fig:supplementary-accuracy-cost-benchmarks}
\end{figure}
\clearpage
\endgroup
\restoregeometry

\clearpage
\begingroup
\fontsize{11}{13.6}\selectfont
\pagestyle{plain}
\renewcommand{\small}{\fontsize{10}{12}\selectfont}
\renewcommand{\footnotesize}{\fontsize{9}{11}\selectfont}
\renewcommand{\scriptsize}{\fontsize{8}{9.5}\selectfont}
\renewcommand{\tablename}{Supplementary Table}
\renewcommand{\thetable}{\arabic{table}}
\renewcommand{\theHtable}{supplementary.\arabic{table}}
\setcounter{table}{0}
\captionsetup[table]{
  font={normalfont,small},
  labelfont=bf,
  justification=justified,
  singlelinecheck=false,
  labelsep=barsep,
  margin=0pt
}

\begin{center}
  {\LARGE\bfseries Supplementary Information\par}
  \vspace{1.5em}
  {\Large\bfseries Physical Reasoning in Large Language Models is Shaped by Human Language\par}
  \vspace{1.2em}
  \textbf{Gregorii A. Semakin}$^{1,2,*}$, \quad
  \textbf{Timofey V. Losev}$^{1}$, \quad
  \textbf{Ilya V. Prolomov}$^{1}$, \\
  \textbf{Stepan N. Ostarkov}$^{1}$, \quad
  \textbf{Igor V. Alabugin}$^{3}$, \quad
  \textbf{Michael G. Medvedev}$^{1,*}$ \\[0.8em]
  $^{1}$N. D. Zelinsky Institute of Organic Chemistry RAS, 119991 Moscow, Russia \\
  $^{2}$National Research University Higher School of Economics, 101000 Moscow, Russia \\
  $^{3}$Department of Chemistry and Biochemistry, Florida State University, Tallahassee, FL 32306, USA \\[0.5em]
  \texttt{\{gregoriisemakin@gmail.com, medvedev.m.g@gmail.com\}}
\end{center}

\clearpage

\begin{table}[H]
  \centering
  \caption{\textbf{Structural and conformational-diversity descriptors for the LLMConfBench molecules.} $N_{\mathrm{conf}}$ is the number of conformers, $N_{\mathrm{rot}}$ is the number of rotatable bonds, and the TFD and heavy-atom RMSD columns summarize pairwise values across all conformer pairs (mean and maximum). Extended Data Fig.~\ref{fig:supplementary-ensemble-diversity}c shows the corresponding per-conformer values relative to the first conformer. Rows follow the molecule identifiers used in the Extended Data Fig.~\ref{fig:supplementary-molecule-performance}b heatmap; chemical structures are depicted in Extended Data Fig.~\ref{fig:supplementary-molecule-performance}a.}
  \label{tab:supplementary-descriptors}
  \small
  \setlength{\tabcolsep}{4pt}
  \renewcommand{\arraystretch}{1.2}
  \begin{tabularx}{\textwidth}{@{}r>{\raggedright\arraybackslash}Xrrrrrr@{}}
    \toprule
    ID & SMILES & $N_{\mathrm{conf}}$ & $N_{\mathrm{rot}}$ & \shortstack{TFD\\mean} & \shortstack{TFD\\max} & \shortstack{RMSD\\mean (\AA)} & \shortstack{RMSD\\max (\AA)} \\
    \midrule
    1 & \smilescode{CC\#CCOC[C@@H](C)O} & 30 & 3 & 0.433 & 0.957 & 1.780 & 2.707 \\
    2 & \smilescode{CCC[C@@H](CO)N1CC1} & 30 & 4 & 0.388 & 0.716 & 1.504 & 2.344 \\
    3 & \smilescode{C[C@@H](CO)[C@@H](O)[C@H]1CN1} & 30 & 3 & 0.415 & 0.756 & 1.743 & 2.911 \\
    4 & \smilescode{O=CNC[C@H](O)[C@H]1CO1} & 30 & 4 & 0.447 & 0.867 & 1.659 & 2.632 \\
    5 & \smilescode{CCOC(=O)[C@@H](C)N} & 30 & 2 & 0.194 & 0.413 & 1.675 & 2.509 \\
    6 & \smilescode{CC[C@@H](O)[C@@H](C)CCO} & 30 & 4 & 0.382 & 0.795 & 1.574 & 2.346 \\
    7 & \smilescode{CC(C)[C@@H](O)C1(O)CC1} & 30 & 2 & 0.151 & 0.350 & 1.840 & 2.764 \\
    8 & \smilescode{O=CC[C@@H]1CO[C@@H]1CO} & 30 & 3 & 0.355 & 0.719 & 1.369 & 2.009 \\
    9 & \smilescode{CCc1ccc(CO)[nH]1} & 30 & 2 & 0.242 & 0.518 & 1.318 & 1.949 \\
    10 & \smilescode{CC(=O)[C@H](C=O)[C@@H](C)O} & 30 & 3 & 0.464 & 0.903 & 1.820 & 2.816 \\
    11 & \smilescode{OCC[C@H]1C[C@@H](O)C1} & 30 & 2 & 0.184 & 0.432 & 1.395 & 2.338 \\
    12 & \smilescode{C[C@@H](O)/C=C/[C@@H](C)OO} & 30 & 3 & 0.197 & 0.385 & 1.618 & 2.545 \\
    13 & \smilescode{N\#CCOCCC1CC1} & 30 & 4 & 0.383 & 0.749 & 1.814 & 2.755 \\
    14 & \smilescode{CCN[C@@H](C\#N)C(N)=O} & 30 & 3 & 0.456 & 0.876 & 1.629 & 2.463 \\
    15 & \smilescode{C\#CC[C@](C)(O)CC=O} & 30 & 3 & 0.463 & 0.731 & 1.717 & 2.773 \\
    16 & \smilescode{C[C@@H](CO)CCNC=O} & 30 & 5 & 0.428 & 0.865 & 1.738 & 2.788 \\
    17 & \smilescode{C[C@@H]1C=CC[C@H](CO)C1} & 30 & 1 & 0.268 & 0.507 & 1.273 & 2.001 \\
    18 & \smilescode{C[C@H]1CCO[C@H]1CCO} & 30 & 2 & 0.235 & 0.610 & 1.387 & 2.316 \\
    19 & \smilescode{O[C@H]1C[C@H]1CCC1CC1} & 30 & 3 & 0.330 & 0.608 & 1.797 & 2.677 \\
    20 & \smilescode{CC(=N)OCCCCO} & 30 & 4 & 0.348 & 0.653 & 1.769 & 2.475 \\
    21 & \smilescode{OCC[C@H]1C[C@H]2OC[C@@H]12} & 30 & 2 & 0.446 & 0.862 & 1.263 & 2.024 \\
    22 & \smilescode{CC(=O)C[C@@]1(O)C[C@H]1O} & 30 & 2 & 0.424 & 0.809 & 1.592 & 2.428 \\
    23 & \smilescode{CC[C@@H](COC)N1CC1} & 30 & 4 & 0.364 & 0.701 & 1.496 & 2.489 \\
    24 & \smilescode{CC(=O)OC1\textbackslash{}C=C(C)/CCC2OC2(C)CC3OC(=O)C(=C)C13} & 30 & 1 & 0.217 & 0.504 & 2.176 & 3.765 \\
    25 & \smilescode{OCC1=CCCCOC1} & 30 & 1 & 0.322 & 0.600 & 1.271 & 1.989 \\
    26 & \smilescode{C[C@H](C=O)COC=N} & 30 & 4 & 0.375 & 0.773 & 1.642 & 2.580 \\
    27 & \smilescode{O=CCCC\#C[C@H]1CN1} & 30 & 2 & 0.274 & 0.447 & 1.645 & 2.674 \\
    \bottomrule
  \end{tabularx}
\end{table}

\clearpage
\begin{table}[H]
  \centering
  \caption{\textbf{Reasoning settings and parameter counts for the cohort.} (T = temperature).}
  \label{tab:supplementary-reasoning}
  \scriptsize
  \setlength{\tabcolsep}{4pt}
  \renewcommand{\arraystretch}{1.6}
  \begin{tabularx}{\textwidth}{@{}>{\raggedright\arraybackslash}p{0.25\textwidth}>{\raggedright\arraybackslash}X>{\raggedright\arraybackslash}p{0.24\textwidth}@{}}
    \toprule
    Model & Reasoning & Parameters \\
    \midrule
    \textbf{GPT-6 Astra} & Effort high; T omitted & --- \\
    \textbf{Claude Opus 5} & Effort high & --- \\
    \textbf{Claude Sonnet 5} & Effort high & --- \\
    \emph{Claude Sonnet 4.5} & Reasoning enabled & --- \\
    \emph{Claude Sonnet 4} & Reasoning enabled & --- \\
    \textbf{GPT-5.6 Sol} & Effort high; T omitted & --- \\
    \textbf{GPT-5} & Effort high; T omitted & --- \\
    \emph{o4-mini} & Reasoning enabled; T omitted & --- \\
    \textbf{Gemini 3.6 Flash} & Effort high & --- \\
    \textbf{Gemini 3 Flash} & Effort high & --- \\
    \emph{Gemini 2.5 Flash} & Reasoning enabled & --- \\
    Gemma 3 27B & No thinking & 27B dense \\
    \textbf{Gemma 4 31B} & Thinking; effort high & 31B dense \\
    \textbf{Kimi K3} & Effort high & 2,800B (104B) MoE \\
    \emph{Kimi K2 Thinking} & Reasoning enabled & 1,100B (32B) MoE \\
    \textbf{Qwen 3.8 27B} & Thinking; effort high & 27.8B dense \\
    \emph{Qwen 3.7 Max} & Reasoning enabled & --- \\
    \textbf{Qwen 3.6 27B} & Thinking; effort high & 27.8B dense \\
    \textbf{Qwen 3.5 27B} & Thinking; effort high & 27.8B dense \\
    \emph{Qwen 3.5 Plus 2026-04-20} & Reasoning enabled & 397B (17B) MoE \\
    Qwen 2.5 32B & No thinking & 32.8B dense \\
    \textbf{Qwen 3 32B} & Thinking; effort high & 32.8B dense \\
    \textbf{Qwen 3.5 35B A3B} & Thinking; effort high & 35B (3B) MoE \\
    \textbf{Qwen 3.6 35B A3B} & Thinking; effort high & 35B (3B) MoE \\
    \textbf{DeepSeek V4 Flash 0731} & Effort high & 284B (13B) MoE \\
    \emph{DeepSeek V3.2} & Reasoning enabled & 671B (37B) MoE \\
    \textbf{DeepSeek-R1 32B} & Thinking; effort high & 32.8B dense \\
    \textbf{Muse Glimmer 30B} & Thinking; effort high & 30B dense \\
    \bottomrule
  \end{tabularx}
\end{table}

\clearpage
\begin{table}[H]
  \centering
  \caption{\textbf{Inference settings shared across the benchmark run.}}
  \label{tab:supplementary-inference-settings}
  \small
  \setlength{\tabcolsep}{5pt}
  \renewcommand{\arraystretch}{1.2}
  \begin{tabularx}{\textwidth}{@{}>{\raggedright\arraybackslash}p{0.30\textwidth}>{\raggedright\arraybackslash}X@{}}
    \toprule
    Setting & Value \\
    \midrule
    System prompt & You are an expert in computational chemistry. \\
    User prompt & Geometry-only XYZ bundle; no SMILES strings or energies \\
    Temperature & 1.0, except where unsupported by the provider schema \\
    Top-p, top-k, and random seed & Not set; provider defaults \\
    Maximum output tokens & Not set; provider default \\
    Tools and web search & Disabled \\
    Enforced JSON response mode & Disabled; JSON was parsed from the free response \\
    \bottomrule
  \end{tabularx}
\end{table}

\clearpage
\begin{table}[H]
  \centering
  \caption{\textbf{Performance of the evaluated models.} Models and force fields are ordered by mean CRA across all reliable pairs (3,802 pairs). For each molecule, scores from the three prompt presentations are averaged before aggregation across the 27 molecules---except for Qwen 2.5 32B and Qwen 3 32B, which returned 25 and 26 valid answers, respectively; SD reports variation of per-molecule means.}
  \label{tab:model-performance-all}
  \scriptsize
  \setlength{\tabcolsep}{4pt}
  \renewcommand{\arraystretch}{1.3}
  \begin{tabularx}{\textwidth}{@{}>{\raggedright\arraybackslash}X>{\raggedright\arraybackslash}p{0.19\textwidth}rrrr@{}}
    \toprule
    \shortstack{Model or\\force field} & \shortstack{Year and month\\of introduction} & \shortstack{Mean\\CRA} & \shortstack{Median\\CRA} & \shortstack{SD\\of CRA} & \shortstack{Count $n$\\(out of 27)} \\
    \midrule
    MMFF94 & April 1996 & 0.93 & 0.977 & 0.106 & 27 \\
    GFN-FF & April 2020 & 0.72 & 0.846 & 0.322 & 27 \\
    \textbf{GPT-6 Astra} & September 2026 & 0.67 & 0.797 & 0.363 & 27 \\
    \textbf{Claude Opus 5} & July 2026 & 0.61 & 0.624 & 0.291 & 27 \\
    \textbf{Gemini 3.6 Flash} & July 2026 & 0.43 & 0.544 & 0.366 & 27 \\
    \textbf{GPT-5.6 Sol} & July 2026 & 0.40 & 0.467 & 0.383 & 27 \\
    \textbf{Kimi K3} & July 2026 & 0.38 & 0.390 & 0.370 & 27 \\
    \textbf{Claude Sonnet 5} & June 2026 & 0.37 & 0.440 & 0.479 & 27 \\
    \textbf{DeepSeek V4 Flash 0731} & July 2026 & 0.33 & 0.422 & 0.414 & 27 \\
    \textbf{Gemini 3 Flash} & December 2025 & 0.32 & 0.379 & 0.430 & 27 \\
    \emph{Qwen 3.7 Max} & May 2026 & 0.32 & 0.274 & 0.328 & 27 \\
    \textbf{Qwen 3.8 27B} & August 2026 & 0.31 & 0.397 & 0.358 & 27 \\
    \textbf{Muse Glimmer 30B} & August 2026 & 0.30 & 0.291 & 0.395 & 27 \\
    \textbf{GPT-5} & August 2025 & 0.30 & 0.392 & 0.506 & 27 \\
    \emph{Kimi K2 Thinking} & November 2025 & 0.27 & 0.293 & 0.307 & 27 \\
    \textbf{Gemma 4 31B} & April 2026 & 0.27 & 0.378 & 0.407 & 27 \\
    UFF & December 1992 & 0.23 & 0.279 & 0.619 & 27 \\
    \textbf{Qwen 3.5 27B} & February 2026 & 0.19 & 0.251 & 0.341 & 27 \\
    Qwen 3.5 Plus & --- & 0.18 & 0.216 & 0.342 & 27 \\
    HBC & --- & 0.16 & 0.000 & 0.237 & 27 \\
    \emph{DeepSeek V3.2} & December 2025 & 0.14 & 0.137 & 0.333 & 27 \\
    \emph{Claude Sonnet 4.5} & September 2025 & 0.14 & 0.127 & 0.301 & 27 \\
    \textbf{Qwen 3.6 27B} & April 2026 & 0.12 & 0.057 & 0.245 & 27 \\
    O4 Mini & --- & 0.08 & 0.021 & 0.267 & 27 \\
    \textbf{Qwen 3.5 35B A3B} & February 2026 & 0.06 & 0.054 & 0.157 & 27 \\
    \emph{Claude Sonnet 4} & May 2025 & 0.06 & 0.045 & 0.187 & 27 \\
    \emph{Gemini 2.5 Flash} & June 2025 & 0.05 & 0.026 & 0.155 & 27 \\
    \textbf{Qwen 3.6 35B A3B} & April 2026 & 0.04 & 0.053 & 0.210 & 27 \\
    Gemma 3 27B & March 2025 & 0.03 & 0.020 & 0.194 & 27 \\
    \textbf{DeepSeek-R1 32B} & January 2025 & 0.01 & -0.005 & 0.122 & 27 \\
    \textbf{Qwen 3 32B} & April 2025 & 0.00 & 0.004 & 0.190 & 26 \\
    Qwen 2.5 32B & September 2024 & 0.00 & -0.022 & 0.154 & 25 \\
    \bottomrule
  \end{tabularx}
\end{table}

\clearpage
\begin{table}[H]
  \centering
  \caption{\textbf{Performance of the evaluated models.} Models and force fields are ordered by mean CRA across fair pairs (2,997 pairs). For each molecule, scores from the three prompt presentations are averaged before aggregation across the 27 molecules---except for Qwen 2.5 32B and Qwen 3 32B, which returned 25 and 26 valid answers, respectively; SD reports variation of per-molecule means. The hydrogen-bond counting rule (HBC) is at chance on fair pairs by construction, because both conformers of every fair pair have the same number of H-bonds; its CRA on all reliable pairs is 0.16 (Fig.~\ref{fig:benchmark}b).}
  \label{tab:model-performance}
  \scriptsize
  \setlength{\tabcolsep}{4pt}
  \renewcommand{\arraystretch}{1.3}
  \begin{tabularx}{\textwidth}{@{}>{\raggedright\arraybackslash}X>{\raggedright\arraybackslash}p{0.19\textwidth}rrrr@{}}
    \toprule
    \shortstack{Model or\\force field} & \shortstack{Year and month\\of introduction} & \shortstack{Mean\\CRA} & \shortstack{Median\\CRA} & \shortstack{SD\\of CRA} & \shortstack{Count $n$\\(out of 27)} \\
    \midrule
    MMFF94 & April 1996 & 0.93 & 0.984 & 0.112 & 27 \\
    GFN-FF & April 2020 & 0.66 & 0.789 & 0.385 & 27 \\
    \textbf{GPT-6 Astra} & September 2026 & 0.66 & 0.793 & 0.380 & 27 \\
    \textbf{Claude Opus 5} & July 2026 & 0.55 & 0.574 & 0.317 & 27 \\
    UFF & December 1992 & 0.42 & 0.615 & 0.592 & 27 \\
    \textbf{Gemini 3.6 Flash} & July 2026 & 0.34 & 0.395 & 0.364 & 27 \\
    \textbf{GPT-5.6 Sol} & July 2026 & 0.30 & 0.293 & 0.379 & 27 \\
    \textbf{Kimi K3} & July 2026 & 0.27 & 0.300 & 0.392 & 27 \\
    \textbf{Claude Sonnet 5} & June 2026 & 0.25 & 0.298 & 0.486 & 27 \\
    \textbf{DeepSeek V4 Flash 0731} & July 2026 & 0.23 & 0.299 & 0.371 & 27 \\
    \textbf{Gemini 3 Flash} & December 2025 & 0.23 & 0.268 & 0.399 & 27 \\
    \emph{Qwen 3.7 Max} & May 2026 & 0.21 & 0.140 & 0.295 & 27 \\
    \textbf{Qwen 3.8 27B} & August 2026 & 0.20 & 0.171 & 0.349 & 27 \\
    \textbf{Muse Glimmer 30B} & August 2026 & 0.20 & 0.208 & 0.416 & 27 \\
    \textbf{GPT-5} & August 2025 & 0.19 & 0.158 & 0.482 & 27 \\
    \textbf{Gemma 4 31B} & April 2026 & 0.17 & 0.148 & 0.378 & 27 \\
    \emph{Kimi K2 Thinking} & November 2025 & 0.17 & 0.181 & 0.292 & 27 \\
    \textbf{Qwen 3.5 27B} & February 2026 & 0.15 & 0.172 & 0.345 & 27 \\
    \emph{Qwen 3.5 Plus 2026-04-20} & April 2026 & 0.08 & 0.103 & 0.326 & 27 \\
    \emph{Claude Sonnet 4.5} & September 2025 & 0.078 & 0.071 & 0.289 & 27 \\
    \emph{DeepSeek V3.2} & December 2025 & 0.077 & 0.075 & 0.297 & 27 \\
    \textbf{Qwen 3.6 27B} & April 2026 & 0.068 & 0.052 & 0.262 & 27 \\
    \emph{o4-mini} & April 2025 & 0.062 & 0.021 & 0.261 & 27 \\
    Gemma 3 27B & March 2025 & 0.035 & 0.014 & 0.186 & 27 \\
    \textbf{Qwen 3.5 35B A3B} & February 2026 & 0.031 & 0.045 & 0.144 & 27 \\
    \emph{Gemini 2.5 Flash} & June 2025 & 0.029 & 0.024 & 0.128 & 27 \\
    \emph{Claude Sonnet 4} & May 2025 & 0.024 & 0.042 & 0.188 & 27 \\
    \textbf{Qwen 3.6 35B A3B} & April 2026 & 0.018 & 0.029 & 0.237 & 27 \\
    Qwen 2.5 32B & September 2024 & 0.010 & -0.019 & 0.164 & 25 \\
    \textbf{DeepSeek-R1 32B} & January 2025 & 0.004 & 0.024 & 0.145 & 27 \\
    HBC (H-bond counting rule) & --- & 0.00 & 0.00 & 0.00 & 27 \\
    \textbf{Qwen 3 32B} & April 2025 & -0.012 & 0.028 & 0.182 & 26 \\
    \bottomrule
  \end{tabularx}
\end{table}

\clearpage
\begin{table}[H]
  \centering
  \caption{\textbf{Paired molecule-level comparisons with UFF for all evaluated models.} Comparisons use the same 27 benchmark molecules (across all 3,802 reliable pairs), except for Qwen 3 32B and Qwen 2.5 32B, for which 26 and 25 molecules, respectively, were available. Differences are calculated as $d_i=\tau_i^{\mathrm{LLM}}-\tau_i^{\mathrm{UFF}}$ at the 2.0~kcal~mol$^{-1}$ energy-gap threshold. The interval is the 95\% Student-$t$ confidence interval for the mean; $p_{\mathrm{perm}}$ and $p_{\mathrm{W}}$ are the two-sided paired permutation-test and Wilcoxon signed-rank $p$-values, respectively.}
  \label{tab:supplementary-uff-comparisons}
  \small
  \setlength{\tabcolsep}{6pt}
  \renewcommand{\arraystretch}{1.1}
  \begin{tabularx}{\textwidth}{@{}>{\raggedright\arraybackslash}Xrrrrrr@{}}
    \toprule
    Model & Mean $d_i$ & Median $d_i$ & \shortstack{CI\\lower} & \shortstack{CI\\upper} & $p_{\mathrm{perm}}$ & $p_{\mathrm{W}}$ \\
    \midrule
    \textbf{GPT-6 Astra} & 0.444 & 0.446 & 0.099 & 0.789 & 0.015 & 0.018 \\
    \textbf{Claude Opus 5} & 0.380 & 0.211 & 0.060 & 0.700 & 0.023 & 0.041 \\
    \textbf{Gemini 3.6 Flash} & 0.202 & 0.328 & -0.150 & 0.554 & 0.248 & 0.361 \\
    \textbf{GPT-5.6 Sol} & 0.168 & 0.387 & -0.156 & 0.492 & 0.296 & 0.336 \\
    \textbf{Kimi K3} & 0.151 & 0.133 & -0.141 & 0.442 & 0.296 & 0.441 \\
    \textbf{Claude Sonnet 5} & 0.143 & 0.387 & -0.183 & 0.468 & 0.382 & 0.397 \\
    \textbf{DeepSeek V4 Flash 0731} & 0.101 & 0.000 & -0.221 & 0.423 & 0.522 & 0.620 \\
    \textbf{Gemini 3 Flash} & 0.087 & 0.071 & -0.263 & 0.437 & 0.609 & 0.714 \\
    \emph{Qwen 3.7 Max} & 0.087 & 0.164 & -0.231 & 0.404 & 0.580 & 0.696 \\
    \textbf{Qwen 3.8 27B} & 0.085 & 0.123 & -0.226 & 0.395 & 0.573 & 0.714 \\
    \textbf{Muse Glimmer 30B} & 0.075 & 0.154 & -0.251 & 0.401 & 0.636 & 0.694 \\
    \textbf{GPT-5} & 0.067 & 0.399 & -0.296 & 0.430 & 0.703 & 0.732 \\
    \emph{Kimi K2 Thinking} & 0.046 & 0.219 & -0.250 & 0.343 & 0.751 & 0.897 \\
    \textbf{Gemma 4 31B} & 0.039 & 0.147 & -0.274 & 0.352 & 0.801 & 0.750 \\
    \textbf{Qwen 3.5 27B} & -0.036 & 0.139 & -0.323 & 0.252 & 0.798 & 0.804 \\
    \emph{Qwen 3.5 Plus 2026-04-20} & -0.048 & -0.064 & -0.346 & 0.250 & 0.738 & 0.696 \\
    \emph{DeepSeek V3.2} & -0.085 & -0.155 & -0.399 & 0.230 & 0.580 & 0.594 \\
    \emph{Claude Sonnet 4.5} & -0.088 & -0.120 & -0.367 & 0.190 & 0.513 & 0.578 \\
    \textbf{Qwen 3.6 27B} & -0.107 & 0.023 & -0.369 & 0.154 & 0.398 & 0.485 \\
    \emph{o4-mini} & -0.144 & -0.067 & -0.415 & 0.126 & 0.278 & 0.313 \\
    \textbf{Qwen 3.5 35B A3B} & -0.170 & -0.215 & -0.435 & 0.096 & 0.201 & 0.166 \\
    \emph{Claude Sonnet 4} & -0.170 & -0.093 & -0.438 & 0.098 & 0.197 & 0.258 \\
    \emph{Gemini 2.5 Flash} & -0.177 & -0.182 & -0.449 & 0.095 & 0.188 & 0.229 \\
    \textbf{Qwen 3.6 35B A3B} & -0.192 & -0.265 & -0.458 & 0.075 & 0.143 & 0.170 \\
    Qwen 2.5 32B & -0.192 & -0.143 & -0.467 & 0.083 & 0.158 & 0.134 \\
    Gemma 3 27B & -0.199 & -0.252 & -0.477 & 0.079 & 0.149 & 0.123 \\
    \textbf{Qwen 3 32B} & -0.208 & -0.359 & -0.501 & 0.086 & 0.154 & 0.143 \\
    \textbf{DeepSeek-R1 32B} & -0.223 & -0.323 & -0.478 & 0.032 & 0.083 & 0.077 \\
    \bottomrule
  \end{tabularx}
\end{table}

\clearpage
\clearpage
\begin{table}[H]
  \centering
  \caption{\textbf{Model identities, access routes, and dates for the frozen cohort.} Release and access dates are shown in that order. Bold font denotes models run with high reasoning effort; italic font denotes models with reasoning enabled at a non-high effort setting; entries without emphasis are models for which reasoning was not applicable.}
  \label{tab:supplementary-model-identities}
  \scriptsize
  \setlength{\tabcolsep}{3pt}
  \renewcommand{\arraystretch}{1.5}
  \begin{tabularx}{\textwidth}{@{}>{\raggedright\arraybackslash}p{0.20\textwidth}>{\raggedright\arraybackslash}X>{\raggedright\arraybackslash}p{0.22\textwidth}>{\raggedright\arraybackslash}p{0.25\textwidth}@{}}
    \toprule
    Model & Exact identifier & Developer / provider & Released / accessed \\
    \midrule
    \textbf{GPT-6 Astra} & \url{openai/gpt-6-astra} & OpenAI / OpenRouter & 2026-09-03 / 2026-09-10--11 \\
    \textbf{Claude Opus 5} & \url{anthropic/claude-opus-5} & Anthropic / OpenRouter & 2026-07-24 / 2026-07-31--08-01 \\
    \textbf{Claude Sonnet 5} & \url{anthropic/claude-sonnet-5} & Anthropic / OpenRouter & 2026-06-30 / 2026-08-03 \\
    \emph{Claude Sonnet 4.5} & \url{anthropic/claude-sonnet-4.5} & Anthropic / OpenRouter & 2025-09-29 / 2026-08-18 \\
    \emph{Claude Sonnet 4} & \url{anthropic/claude-sonnet-4} & Anthropic / OpenRouter & 2025-05-22 / 2026-08-18 \\
    \textbf{GPT-5.6 Sol} & \url{openai/gpt-5.6-sol} & OpenAI / OpenRouter & 2026-07-09 / 2026-07-31 \\
    \textbf{GPT-5} & \url{openai/gpt-5} & OpenAI / OpenRouter & 2025-08-07 / 2026-08-17 \\
    \emph{o4-mini} & \url{openai/o4-mini} & OpenAI / OpenRouter & 2025-04-16 / 2026-08-18 \\
    \textbf{Gemini 3.6 Flash} & \url{google/gemini-3.6-flash} & Google / OpenRouter & 2026-07-21 / 2026-07-31--08-03 \\
    \textbf{Gemini 3 Flash} & \url{google/gemini-3-flash-preview} & Google / OpenRouter & 2025-12-17 / 2026-08-18 \\
    \emph{Gemini 2.5 Flash} & \url{google/gemini-2.5-flash} & Google / OpenRouter & 2025-06-17 / 2026-08-18 \\
    Gemma 3 27B & \url{gemma3:27b} & Google / Ollama & 2025-03-12 / 2026-08-17 \\
    \textbf{Gemma 4 31B} & \url{gemma4:31b} & Google / Ollama & 2026-04-02 / 2026-08-17 \\
    \textbf{Kimi K3} & \url{moonshotai/kimi-k3} & Moonshot AI / OpenRouter & 2026-07-16 / 2026-07-31 \\
    \emph{Kimi K2 Thinking} & \url{moonshotai/kimi-k2-thinking} & Moonshot AI / OpenRouter & 2025-11-06 / 2026-08-16 \\
    \textbf{Qwen 3.8 27B} & \url{qwen3.8:27b} & Alibaba (Qwen) / Ollama & 2026-08-14 / 2026-08-19--20 \\
    \emph{Qwen 3.7 Max} & \url{qwen/qwen3.7-max} & Alibaba (Qwen) / OpenRouter & 2026-05-21 / 2026-07-31 \\
    \textbf{Qwen 3.6 27B} & \url{qwen3.6:27b} & Alibaba (Qwen) / Ollama & 2026-04-21 / 2026-08-16 \\
    \textbf{Qwen 3.5 27B} & \url{qwen3.5:27b} & Alibaba (Qwen) / Ollama & 2026-02-24 / 2026-08-17--18 \\
    \emph{Qwen 3.5 Plus 2026-04-20} & \url{qwen/qwen3.5-plus-20260420} & Alibaba (Qwen) / OpenRouter & 2026-04-20 / 2026-07-31 \\
    Qwen 2.5 32B & \url{qwen2.5:32b} & Alibaba (Qwen) / Ollama & 2024-09-19 / 2026-08-15 \\
    \textbf{Qwen 3 32B} & \url{qwen3:32b} & Alibaba (Qwen) / Ollama & 2025-04-28 / 2026-08-16 \\
    \textbf{Qwen 3.5 35B A3B} & \url{qwen3.5:35b-a3b} & Alibaba (Qwen) / Ollama & 2026-02-24 / 2026-08-16 \\
    \textbf{Qwen 3.6 35B A3B} & \url{qwen3.6:35b-a3b} & Alibaba (Qwen) / Ollama & 2026-04-14 / 2026-08-17 \\
    \textbf{DeepSeek V4 Flash 0731} & \url{deepseek/deepseek-v4-flash-0731} & DeepSeek / OpenRouter & 2026-07-31 / 2026-08-03 \\
    \emph{DeepSeek V3.2} & \url{deepseek/deepseek-v3.2} & DeepSeek / OpenRouter & 2025-12-01 / 2026-08-17 \\
    \textbf{DeepSeek-R1 32B} & \url{deepseek-r1:32b} & DeepSeek / Ollama & 2025-01-20 / 2026-08-15 \\
    \textbf{Muse Glimmer 30B} & \url{muse-glimmer:30b} & Meta (Muse) / Ollama & 2026-08-10 / 2026-08-18 \\
    \bottomrule
  \end{tabularx}
\end{table}

\endgroup
\end{document}